\documentclass[journal,twocolumn,10pt,final,letterpaper]{IEEEtran}

\usepackage{cite}
\usepackage{eucal}

\ifCLASSINFOpdf
   \usepackage[pdftex]{graphicx}
		\graphicspath{{../figures/}}
		\DeclareGraphicsExtensions{.pdf,.jpeg,.png}
\else
	\usepackage[dvips]{graphicx}
   \graphicspath{{../figures/}}
   \DeclareGraphicsExtensions{.eps}
\fi

\usepackage[cmex10]{amsmath}
\usepackage{amssymb,amsthm,dsfont}
\usepackage{mathtools}

\newtheorem{remark}{Remark}
\newtheorem{lemma}{Lemma}
\newtheorem{theorem}{Theorem}
\newtheorem{corollary}{Corollary}
\newtheorem{definition}{Definition}

\newtheorem{proposition}{Proposition}

\newcommand{\x}{\mathsf{x}}
\newcommand{\h}{\mathsf{h}}

\newcommand{\y}{\mathsf{y}}
\newcommand{\I}{\mathsf{I}}
\newcommand{\hypergeom}[4]{{}_{2}\hspace{-.02cm}F_{\hspace{-.02cm}1}\hspace{-.1cm}\left(#1,#2,#3;#4\right)}
\newcommand{\hypergeomexp}[4]{{}_{2}\hspace{-.02cm}F_{\hspace{-.02cm}1}\hspace{-.04cm}\left(#1,#2,#3;#4\right)}
\newcommand{\cp}{\mathtt{P}_{\text{\textnormal{c}}}}

\newcommand{\cpia}{\mathtt{P}_{\text{\textnormal{c}}}^{\text{\textnormal{IA}}}}
\newcommand{\cpib}{\mathtt{P}_{\text{\textnormal{c}}}^{\text{\textnormal{IB}}}}
\newcommand{\cpsiso}{\mathtt{P}_{\textnormal{c}}^{\textnormal{SISO}}}
\newcommand{\cpnc}{\mathtt{P}^{\text{\textnormal{IA}}}_{\text{\textnormal{c,NC}}}}
\newcommand{\cpfc}{\mathtt{P}^{\text{\textnormal{IA}}}_{\text{\textnormal{c,FC}}}}

\newcommand{\sinr}{\mathtt{SINR}}

\newcommand{\var}{\text{\textnormal{Var}}}
\newcommand{\cov}{\text{\textnormal{Cov}}}
\newcommand{\lap}{\mathcal{L}}

\renewcommand{\th}{\text{-th}}

\newcommand{\cpsc}{\mathtt{P}_{\textnormal{c}}^{\textnormal{SC}}}

\newcommand{\snr}{\mathtt{SNR}}
\newcommand{\mathdef}{~\raisebox{-0.03cm}{$\triangleq$}~}

\usepackage{psfrag}
\usepackage{array}
\usepackage{mdwmath}  
\usepackage{mdwtab}
\usepackage{enumerate}
\usepackage{cases}

\let\underbrace\LaTeXunderbrace

\usepackage[caption=false,font=footnotesize,labelfont=sf,textfont=sf]{subfig}

\usepackage{url}

\allowdisplaybreaks

\begin{document}

\title{Analysis of Joint Transmit-Receive Diversity in Downlink MIMO Heterogeneous Cellular Networks}
%\title{}

\author{Ralph Tanbourgi,~\IEEEmembership{Student Member,~IEEE,} Harpreet S. Dhillon,~\IEEEmembership{Member,~IEEE,}\\and
Friedrich K. Jondral,~\IEEEmembership{Senior Member,~IEEE}

\thanks{R.~Tanbourgi and F.~K.~Jondral are with the Communications Engineering Lab (CEL), Karlsruhe Institute of Technology (KIT), Germany. Email: \texttt{\{ralph.tanbourgi, friedrich.jondral\}@kit.edu}.% This work was partially supported by the German Research Foundation (DFG) within the Priority Program 1397 "COIN" under grant No. JO258/21-2.
}
\thanks{H. S. Dhillon is with the Wireless@VT, Department of ECE, Virgina Tech, Blacksburg, VA, USA. Email: \texttt{hdhillon@vt.edu}.}

\thanks{This work will be presented in part at the IEEE International Conference on Communications 2015, London, UK\cite{tandhi14_3}.}
%\thanks{J. G. Andrews is with the Wireless and Networking Communications Group (WNCG), The University of Texas at Austin, TX, USA. Email: \texttt{jandrews@ece.utexas.edu}.}
%\thanks{Date modified \today.}
}

\maketitle

\begin{abstract}
We study multiple-input multiple-output (MIMO) based downlink heterogeneous cellular network (HetNets) with joint transmit-receive diversity using orthogonal space-time block codes at the base stations (BSs) and maximal-ratio combining (MRC) at the users. MIMO diversity with MRC is especially appealing in cellular networks due to the relatively low hardware complexity at both the BS and user device. Using stochastic geometry, we develop a tractable stochastic model for analyzing such HetNets taking into account the irregular and multi-tier BS deployment. We derive the coverage probability for both interference-blind (IB) and interference-aware (IA) MRC as a function of the relevant tier-specific system parameters such as BS density and power, path loss law, and number of transmit (Tx) antennas. Important insights arising from our analysis for typical HetNets are for instance: (i) IA-MRC becomes less favorable than IB-MRC with Tx diversity due to the smaller interference variance and increased interference correlation across Rx antennas; (ii) ignoring spatial interference correlation significantly overestimates the performance of IA-MRC; (iii) for small number of Rx antennas, selection combining may offer a better complexity-performance trade-off than MRC.
\end{abstract}


\IEEEpeerreviewmaketitle

\section{Introduction}\label{sec:introduction}

%\IEEEPARstart{P}{romising} approaches to address the ever-increasing rate and coverage demands in cellular systems include network densification and addition of more antennas to both base stations (BSs) and user devices, and using multiple-input multiple-output (MIMO) techniques. Network densification  Although considered 

\IEEEPARstart{N}{etwork} densification and multiple-input multiple-output (MIMO) communications are two promising approaches to address the increasing rate and coverage demands in cellular systems\cite{andrews_5G}. Network densification is realized by deploying tiers of low-power base stations (BSs) inside the existing network, e.g., to serve high-traffic areas within macro cells, thereby rendering the network increasingly {\it heterogeneous}. MIMO, on the other hand, can increase link reliability and/or capacity by leveraging the spatial degrees-of-freedom in fading channels. Due to the radical shift associated with heterogeneous cellular networks (HetNets), MIMO and HetNets cannot be analyzed separately; many characteristics unique to HetNets such as multi-tier deployment, limited site-planning, and heterogeneous parametrization clearly influence the channel ``seen'' by a multi-antenna receiver. Understanding this interplay, however, is challenging and makes a comprehensive analysis of MIMO in HetNets difficult. This paper addresses this challenge and develops a tractable model to study the performance of MIMO diversity in HetNets.%\vspace{-.5cm}

\subsection{Motivation and Related Work}\label{sec:motivation}

MIMO techniques can be open-loop or closed-loop based and the latter have been the focus of many works on MIMO cellular networks, see for instance\cite{gesbert07} and the references therein. These works show that closed-loop MIMO schemes can significantly improve performance when channel state information at the transmitter (CSI-T) is available. Reliable CSI-T, however, may not always be available in practice, e.g., in high mobility scenarios\cite{spencer04mag}, and open-loop schemes requiring CSI only at the receiver (CSI-R) have to be used instead. For instance, 3rd Generation Partnership Program (3GPP) Long Term Evolution (LTE) supports different open-loop modes, e.g., transmission mode 2 uses a space-frequency block code (SFBC) for transmit (Tx) diversity over two or four Tx antennas\cite{3gpp_36213_mimo}. On the mobile receiver side, space and complexity limitations typically preclude the use of many receive (Rx) antennas---often not exceeding two antennas---and allow only for simple linear combining schemes. One such combining scheme is maximal-ratio combining (MRC) \cite{brennan59}, which offers a good trade-off between performance and complexity, and is therefore ubiquitously found in multi-antenna consumer devices. Especially in the context of MIMO communications, MRC may sometimes be even more appealing than interference-canceling receivers, since the latter require accurate knowledge of the other-cell interference channels, which is harder to realize when multiple Tx antennas are active\cite{tse05}.

Two types of MRC exist that differ in the way interference due to concurrent transmissions is treated, namely {\it interference-blind} (IB) and {\it interference-aware} (IA) MRC. The former, and more popular, ignores the interference at all. The combiner coefficients then follow from the well-known channel matched-filter approach in this case\cite{tse05}. IA-MRC, in contrast, takes the interference power into account. More specifically, the (possibly unequal) interference power experienced at each Rx antenna in one block/frame is treated as additional Rx noise. Following the original MRC approach from\cite{brennan59}, the combiner then give less weight to branches with poor reception quality, i.e., with strong interference and/or adverse fading states. Estimating the per-antenna interference power can be done within the channel estimation phase, e.g., after decoding and removing the pilot symbols sent by the serving BS \cite{gosh10} or by using techniques from\cite{benedict67,pauluzzi00}. Both types of MRC are well-understood for networks with fixed geometry, see for instance\cite{cui99,aalo00,cui04}, and recently also for networks with dynamic/varying geometry\cite{hunter08,chopra12,tandhi14_1,tandhi14_2}. IA-MRC was also studied for single-tier single-Tx antenna cellular networks in \cite{tanbourgi14_1}.

In the context of downlink HetNets, MIMO diversity with MRC is not yet well-understood, since prior works do not directly apply due to the specific nature of the interference governing HetNets. An interesting question, for instance, is whether the gain of IA-MRC over IB-MRC justifies the slightly higher complexity in a typical MIMO HetNet setting, and how this trade-off varies with the number of Tx and Rx antennas. %? And how does spatial interference correlation across Rx antennas affect the performance of MRC? 
Certainly, extensive system-level simulations can only partly help in addressing the above challenge as they usually offer only limited insights. As a viable alternative approach to simulations, spatial modeling using stochastic geometry\cite{stoyan95} has gained much attention recently, see for instance\cite{andrews11,jo12,dhillon12,zhang14,haenggi12_1,Haenggi14twc,govindasamy14,direnzo14,direnzo15,chang14,dhillon13_mimo,gupta14} and the references therein. In the context of MIMO cellular networks, \cite{direnzo14} analyzed the average symbol error probability (ASEP) in multi-antenna single-tier cellular networks with spatial-multiplexing, where it is found that Rx diversity can significantly improve performance. In \cite{direnzo15}, a unifying framework using the Equivalent-in-Distribution approach was presented, which studies the ASEP of MIMO diversity with IB-MRC in single-tier networks. The energy efficiency of small-cell multiple-input single-output (MISO) cellular system with maximal-ratio transmission in the downlink was analyzed in \cite{chang14}. MIMO HetNets with different kinds of CSI-T based MIMO schemes were analyzed in \cite{gupta14} with load balancing and in \cite{dhillon13_mimo} without. Complementing the above works, the main objective of this paper is to derive a tractable model and to conduct a meaningful analysis in order to obtain a better understanding of MIMO diversity with IA/IB-MRC in HetNets.

\subsection{Contributions, Outcomes, and Organization}\label{sec:contributions}
In this work, we study the coverage performance of downlink HetNets with MIMO diversity, where users employ IB-MRC or IA-MRC. Our main contributions are outlined below.

{\bf Analytical model:} In Section~\ref{sec:model}, we develop a tractable model for a downlink $K$-tier HetNet employing MIMO diversity with orthogonal space-time block codes (OSTBCs) and IB-MRC/IA-MRC. To reflect the irregular and multi-tier deployment of BSs in practice, we use a Poisson point process (PPP) to model the BS locations in each tier. The model captures relevant tier-specific parameters, such as BS density and Tx power, path loss exponent, and number of Tx antennas. We derive the coverage probability for both types of MRC in Section~\ref{sec:cp_analysis}. For IA-MRC, we focus on the case with two Rx antennas. The theoretical expressions can be evaluated using standard numerical software and can be further simplified analytically in certain cases.

{\bf Second-order statistics of HetNet interference:} In Section~\ref{sec:int_stats}, we study the interference power (hereafter, simply {\it interference}) dynamics at a multi-antenna receiver in HetNets, thereby complementing earlier work, which focused on Aloha-based ad hoc networks \cite{haenggi12_1}. Interference dynamics affect the performance of IA diversity combining schemes such as IA-MRC. Our analysis shows that the interference variance measured at a typical user is tier-independent for equal path loss exponents and the same the number of Tx antennas across tiers. In direct comparison with the literature, our analysis indicates that the interference variance is smaller in HetNets than in Aloha-based ad hoc networks, where interferers can be much closer to a receiver. Moreover, the gains of IA diversity combining are expected to decrease when more Tx antennas are active (Tx diversity) as interference variance then becomes smaller. Interestingly, the interference correlation coefficient across Rx antennas is independent from the tier with which this user associates when the number of Tx antennas is equal in each tier. In this case, the correlation becomes entirely tier-independent and increases with the number of active Tx antennas. In line with the effect of decreasing interference variance explained above, the gains of IA diversity combining are expected to decrease when more Tx antennas are active due to the higher interference correlation across Rx antennas.

{\bf Design Insights:} In Section~\ref{sec:design}, we discuss the theoretical results using numerical examples. In a typical three-tier MIMO scenario with IB-MRC at the receivers, the gain of doubling the number of Rx antennas is roughly 2.5\,dB at operating points of practical relevance. For IA-MRC, this gain is roughly 3.6\,dB. Adding more Tx antennas is beneficial only at large coverage probabilities. The gain of IA-MRC over IB-MRC decreases with the number of Tx antennas due the higher interference correlation across Rx antennas resulting from the interference-smoothing effect of Tx diversity. The relative coverage probability gain of $1\times2$ single-input multiple-output (SIMO) over SISO transmission in the practical regime is between 12\%--66\% for IB-MRC, while an additional improvement of only 1\%--3\% is obtained by IA-MRC. Although interference power estimation needed in IA-MRC can be realized with affordable complexity, the outcome of this comparison suggests that IA-MRC is less favorable than IB-MRC in MIMO HetNets with Tx diversity.

Spatial interference correlation across Rx antennas, caused by the common locations of interfering BSs, influences the performance of IA-MRC and should not be ignored in the analysis; ignoring it significantly overestimates the true performance. In contrast, assuming full correlation underestimates the true performance only slightly. Moreover, it is shown that assuming full correlation in IA-MRC is equivalent to IB-MRC. Interestingly, it does not matter for the diversity order of IA-MRC if one assumes no correlation or full correlation of the interference. 

We derive the coverage probability of selection combining (SC) for SIMO HetNets and compare its performance to MRC. Our results show that the gain of MRC over SC is not overwhelming for small number of Rx antenna. The higher complexity of MRC might thus outweigh the gain over SC. Moreover, the performance-complexity trade-off between MRC and SC in HetNets may differ significantly from the interference-free case.

{\it Notation:} We use sans-serif-style letters ($\mathsf{z}$) for denoting random variables and serif-style letters ($z$) for denoting their realizations or variables. We define $(z)^{+}\mathdef\max\{0,z\}$.

\section{System Model}\label{sec:model}

\subsection{Network Geometry and User Association}

\begin{figure}[t]
	\centering
	\includegraphics[width=.47\textwidth]{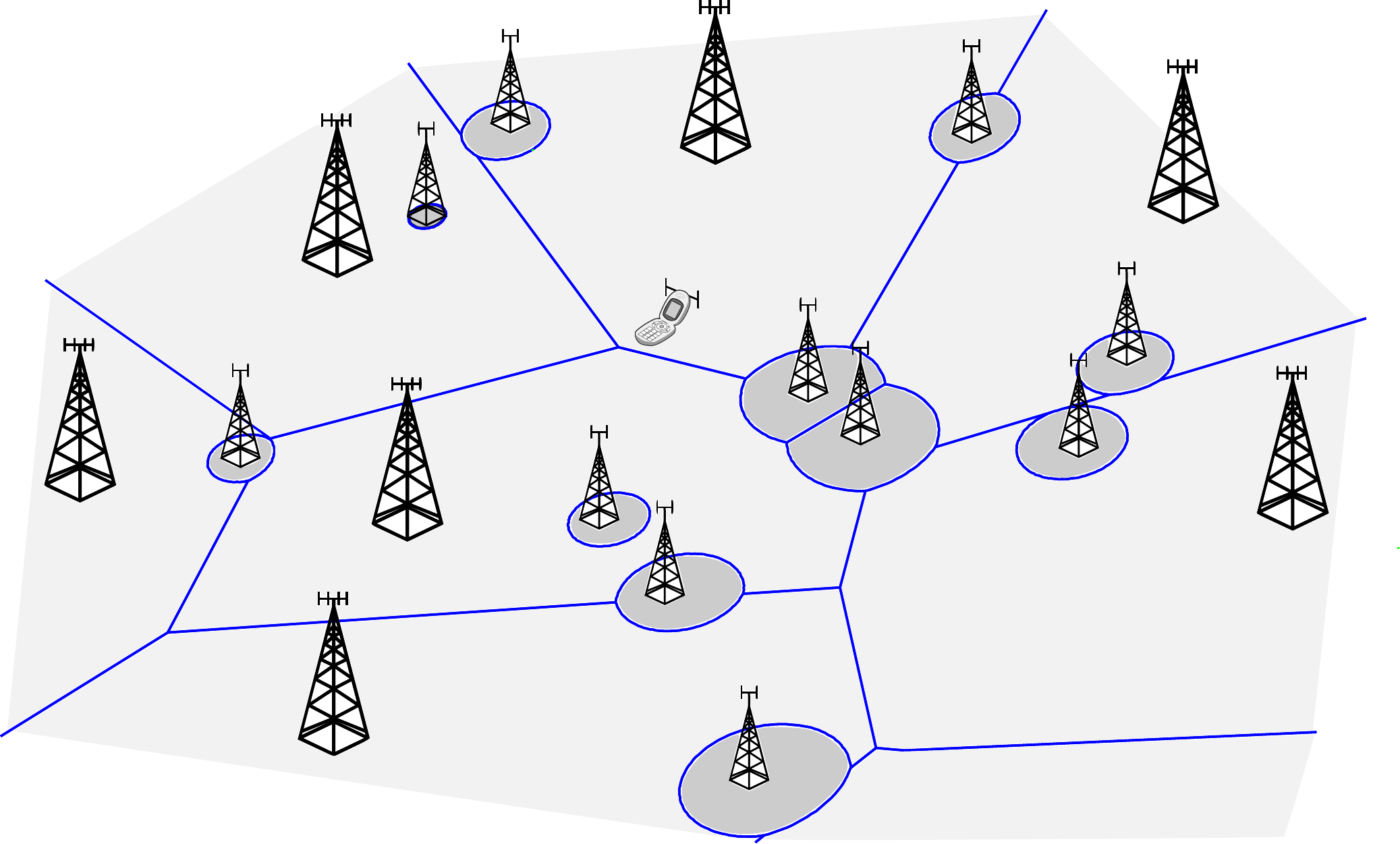}
	\caption{Example: HetNet with $K=2$. Tier-1 BSs have 4 antennas Tier-2 BSs have 2 antennas. Typical user has 2 antennas.}
	\label{fig:illustration}
\end{figure}
We consider a $K$-tier HetNet in the downlink with BSs irregularly scattered in the plane, see Fig.~\ref{fig:illustration}. We model the irregular BS locations in tier $k$ by an independent stationary planar PPP $\Phi_k$ with density $\lambda_k$ and denote by $\Phi\mathdef\cup_{k=1}^{K}\Phi_k$ the entire set of BSs. The spatial Poisson model is widely-accepted for analyzing (multi-tier) cellular networks\cite{andrews11,jo12, dhillon12}, and recently also MIMO HetNets\cite{dhillon13_mimo,gupta14}. All BSs in tier $k$ transmit an OSTBC using $M_k$ Tx antennas. Similarly, we assume that mobile receivers (users) are equipped with $N$ co-located antennas. The users are independently distributed on the plane according to some stationary point process. By Slivnyak's theorem \cite{stoyan95} and due to the stationarity of $\Phi$, we can focus the analysis on a \textit{typical} user located at the origin $o\in\mathbb{R}^2$. BSs in tier $k$ transmit with total power $P_k$, which is equally divided across all active Tx antennas. At the typical user, the long-term received power from a tier $k$ BS located at $\x_i\in\Phi_k$ is thus $P_k\|\x_i\|^{-\alpha_k}$, where $\|\cdot\|^{\alpha_k}$ is the distance-dependent path loss with path loss exponent $\alpha_k>2$. We assume independent and identically distributed frequency-flat Rayleigh fading. Table~\ref{tab:notation} shows the notation used in this work.

Users are assumed to associate with the BS providing the strongest average measured received power, which is a common assumption in cellular systems. Note that this association rule is generally not coverage maximizing in MIMO HetNets with unequal $M_k$ and requires a {\it biased} association rule, see\cite{gupta14} for more details. Including biasing in the model, e.g., using similar techniques as in\cite{jo12,gupta14}, is left for possible future work. For the typical user, the serving BS is hence the one maximizing $P_k\|\x_i\|^{-\alpha_k}$. Without loss of generality, we label the location of this BS by $\x_o$ and denote by $\y\mathdef\|\x_o\|$ its distance to the typical user. For convenience, we define $\Phi^{o}\mathdef\Phi\setminus\{\x_o\}$ (similarly, $\Phi_k^{o}\mathdef\Phi_k\setminus\{\x_o\}$), i.e., the set of interfering BSs. From the association rule discussed above it follows that, given $\y=y$ and that the serving BS is from tier $\ell$, $\Phi^{o}_k$ is a homogeneous PPP on $\mathbb{R}^2\setminus b(0,d_k)$, where $d_k=\hat P_k^{1/\alpha_k}y^{1/\hat \alpha_k}$ with $\hat{P}_{k}\triangleq P_k/P_{\ell}$ and $\hat{\alpha}_k\mathdef\alpha_k/\alpha_{\ell}$.

The following useful lemma gives the probability that a user associates with tier $\ell$ and the conditional probability density function (PDF) of the distance $\y$ to the serving BS.
\begin{lemma}[Association Probability and Distance PDF\cite{jo12}]\label{lem:asso}
A user associates with the $\ell\th$ tier with probability
\begin{IEEEeqnarray}{rCl}
   A_{\ell}=2\pi\lambda_\ell\int_{0}^{\infty}y\,\exp\left(-\pi\sum\limits_{k=1}^{K}\lambda_k \hat{P}_k^{2/\alpha_k}y^{2/\hat{\alpha}_k}\right)\mathrm dy.
\end{IEEEeqnarray}
The PDF of the distance $\mathsf{y}\mathdef\|\x_o\|$ to the serving BS, given that it belongs to tier $\ell$, is
\begin{IEEEeqnarray}{rCl}
   f_{\mathsf{y},\ell}(y)=\frac{2\pi\lambda_\ell y}{A_{\ell}}\exp\left(-\pi\sum\limits_{k=1}^{K}\lambda_k \hat{P}_k^{2/\alpha_k}y^{2/\hat{\alpha}_k}\right),\quad y\geq0.\IEEEeqnarraynumspace
\end{IEEEeqnarray}
\end{lemma}

\begin{table}[!t]
	\renewcommand{\arraystretch}{1.3}
	\caption{Notation used in this work}%\vspace{-.3cm}
	\label{tab:notation}
	\centering
	{\small
	\begin{tabular}{|c|p{6.2cm}|}
		\hline
		\bfseries{Notation} & \hspace{2.4cm}\bfseries{Description}\\
		\hline
		$\Phi_k;\lambda_k$ 		& PPP describing the BS locations in tier $k$; average density of tier-$k$ BSs; $\Phi\mathdef\cup_{k=1}^{K}\Phi_k$\\
		\hline
		$P_k$		&	 BS transmit power in tier $k$\\
		\hline
		$\alpha_k$	&	Path loss exponent in tier $k$\\
		\hline
		$N$; $M_k$	&	Number Rx antennas at the typical user; number of Tx antennas per BS in tier $k$\\
		\hline
		$\mathbf{H}_{i};\mathsf{h}_{i,nm}$	&	 $N\times M_k$ channel matrix between typical user and $i\th$ tier $k$ BS; entries of $\mathbf{H}_{i}$ from $\mathcal{CN}(0,1)$\\
		\hline
		$\sigma^2$	&	Receiver noise (AWGN) power\\
		\hline
		$(M_k,L_k,r_k)$	&	OSTBC with codeword length $L_k$, code rate $r_k$, and $M_k$ Tx antennas\\
		\hline
		$S_k$	& Number of symbols encoded in an $(M_k,L_k,r_k)$-OSTBC; number of active Tx antennas per slot\\
		\hline
		$\mathbf{A}_{nm},\mathbf{B}_{nm}$	&	Dispersion matrices characterizing an OSTBC\\
		\hline
		$\mathsf{I}_{n};\mathsf{I}_{i,\text{eqv}}$	& Interference power at the $n\th$ antenna; interference from $i\th$ BS after diversity combining\\
		\hline
		$\snr_{\ell}(y)$	&	Mean SNR from a serving $\ell\th$ tier BS at distance $y$\\
		\hline	
		$\sinr_{\ell}(y)$	& SINR from a serving $\ell\th$ tier BS at distance $y$\\
		\hline
		$\sinr$	& SINR at the typical user\\
		\hline
		$T$	&	Target SINR\\
		\hline
		$\cp$	&	Coverage probability $\mathbb{P}(\sinr>T)$\\
		\hline
	\end{tabular}
	}%\vspace{-.5cm}
\end{table}

\subsection{OSTBC MIMO Signal Model}\label{sec:model_ostbc}
All BSs in tier $k$ use an $(M_k,L_k,r_k)$-OSTBC, where $L_k\geq1$ is the codeword length and $r_k\in(0,1]$ is the code rate; $L_k$ can be seen as the number of slots needed to convey $S_k=L_k r_k$ symbols using $M_k$ Tx antennas \cite{tarokh99}. For analytical tractability, we shall consider only {\it power-balanced} $(M_k,L_k,r_k)$-OSTBCs, i.e., having the property that exactly $S_k$ symbols are transmitted, or equivalently that exactly $S_k$ Tx antennas are active, in every slot. This allows assigning a constant power load of $P_k/S_k$ to every symbol-antenna pair in every slot. Practical examples of balanced OSTBCs are for instance $(1,1,1)$ (single-antenna), $(2,2,1)$ (Alamouti), $(4,4,1/2)$, and $(4,4,3/4)$, see\cite{shah00,larson08}. We use the notation $\mathbf{v}_{i,\tau}\in\{0,1\}^{M_k}$ to indicate the active Tx antennas of BS $i$ in slot $\tau$, i.e., the $m\th$ entry of $\mathbf{v}_{i,\tau}$ is one if antenna $m$ is active and zero otherwise.

Assume for the moment that the typical user associates with the $\ell\th$ tier. It will then be served by an $(M_\ell,L_\ell,r_\ell)$-OSTBC. The received signal at the typical user in slot $\tau\in\{1,\ldots,L_\ell\}$ can then be expressed by
\begin{IEEEeqnarray}{rCl}
   \mathbf r_\tau = \mathbf{H}_{o}\,\mathbf c_{o,\tau}+\sum\limits_{k=1}^{K}\sum\limits_{\x_i\in\Phi^{o}_k}\mathbf H_{i}\,\mathbf{ c}_{i,\tau}+\mathbf{n}_{\tau},\IEEEeqnarraynumspace\label{eq:mimo_gen1}
\end{IEEEeqnarray}
where
\begin{itemize}
   \item $\mathbf{H}_i\in\mathbb{C}^{N\times M_k}$ is the channel matrix describing the fading from the $i\th$ BS of the $k\th$ tier to the typical user. The entries of $\mathbf{H}_i$, $\mathsf h_{i,nm}$, are $\mathcal{CN}(0,1)$ distributed and typically assumed constant over one codeword.\footnote{When $\mathbf{H}_i$ is time-varying and space-time coding is across multiple channel realizations, technique from \cite{gamal03} can be used to achieve also temporal diversity. Such extensions are outside the scope of this paper.} We assume $\mathbb{E}[\mathsf h_{i,nm}\mathsf h_{j,uv}^{\ast}]=0$ unless $i=j$, $n=u$, and $m=v$.
     
   \item $\mathbf c_{i,\tau}\in\mathbb{C}^{M_k}$ is a vector consisting of the space-time coded symbols of the $i\th$ BS sent from the $S_k\leq M_k$ active Tx antennas in slot $\tau$ and received with average signal strength $\sqrt{\tfrac{P_k}{S_k}}\|\x_i\|^{-\alpha_k/2}$. We assume that $\mathbb{E}\left[\mathbf c_{i,\tau} \mathbf c_{j,\tau}^H\right]=\mathbf{0}_{M_k}$ for all $i\neq j$, where $\mathbf{0}_{M_k}$ is an $M_k\times M_k$ zero matrix. It is reasonable to assume also that $\mathbb{E}[\mathbf c_{i,\tau}]=[0,\ldots,0]^{T}$ and $\mathbb{E}[\mathbf c_{i,\tau} \mathbf c_{i,\tau}^H]=\tfrac{P_k}{S_k} \|\x_i\|^{-\alpha_k}\text{diag}(\mathbf{v}_{i,\tau})$, where $\text{diag}(\mathbf{v}_{i,\tau})$ is a diagonal matrix with entries $\mathbf{v}_{i,\tau}$. The latter assumption follows from the \textit{balanced-power} property of the considered OSTBCs.
   
	\item $\mathbf{n}_{\tau}\in\mathbb{C}^{N}$ is a vector describing the Rx noise with independent $\mathcal{CN}(0,\sigma^2)$ entries.
\end{itemize}
Upon receiving all $L_\ell$ code symbols corresponding to one codeword, the typical user stacks the vectors $\mathbf r_1,\ldots,\mathbf r_{L_\ell}$ to form the new vector
\begin{IEEEeqnarray}{rCl}
   \mathbf{\bar r}&=&
				\begin{bmatrix}
				   \mathbf{H}_{o}\,\mathbf c_{o,1}\\
				   \vdots\\
				   \mathbf{H}_{o}\,\mathbf c_{o,L_\ell}
				\end{bmatrix}
				+\sum\limits_{k=1}^{K}\sum\limits_{\x_i\in\Phi^{o}_k}
				\underbrace{\begin{bmatrix}
				   \mathbf{H}_{i}\,\mathbf c_{i,1}\\
				   \vdots\\
				   \mathbf{H}_{i}\,\mathbf c_{i,L_\ell}
				\end{bmatrix}
				}_{\mathbf{\bar i}_i}
				+
				\begin{bmatrix}
				   \mathbf{n}_{1}\\
				   \vdots\\
				   \mathbf{n}_{L_\ell}
				\end{bmatrix},\label{eq:mimo_gen2}\IEEEeqnarraynumspace
\end{IEEEeqnarray}
where $\mathbf{ \bar i}_i\in\mathbb{C}^{NL_\ell}$ is the interference signal from the $i\th$ BS received over the entire codeword period. With CSI at the receiver, $\mathbf{\bar r}$ is linearly processed/combined to form a decision variable. Two types of MRC are considered, which differ in the amount of CSI required. More specifically, IB-MRC requires knowledge of $\mathbf H_o$, while IA-MRC requires knowledge of $\mathbf H_o$ and of the interference-plus-noise power. The squared Frobenius norm of a matrix $\mathbf{G}\in\mathbb{C}^{U\times V}$ is given by $\|\mathbf G\|_F^2=\sum_{u=1}^{U}\sum_{v=1}^{V}|g_{uv}|^2$. The following lemma will be useful in the later analysis.

\begin{lemma}[Gaussian Matrices]\label{lem:fading_dis} Let $\mathbf G(u)\in\mathbb C^{v\times w}$ have $u\leq vw$ independent $\mathcal{CN}(0,1)$-distributed entries and $vw-u$ zeros. Then, $\|\mathbf G(u)\|_F^2$ is Erlang distributed with shape $u$, rate 1, and cumulative distribution function (CDF) \cite{feller71}
   \begin{IEEEeqnarray}{rCl}
      \mathbb P (\|\mathbf G(u)\|_F^2\leq\theta)=1-e^{-\theta}\sum\limits_{j=0}^{u-1}\frac{\theta^{j}}{j!}.
   \end{IEEEeqnarray}
	and $p\th$ (raw) moment, $p>-u$, 
	\begin{IEEEeqnarray}{rCl}
	  \mathbb{E}\left[\|\mathbf G(u)\|_F^{2p}\right]&=&\frac{1}{\Gamma(u)}\int_{0}^{\infty}z^{p+u-1}e^{-z}\,\mathrm dz=\frac{\Gamma(u+p)}{\Gamma(u)}.\IEEEeqnarraynumspace
	\end{IEEEeqnarray}
\end{lemma}

It is known that the performance of diversity-combining schemes is influenced by the second-order properties of the interference, as reported in\cite{haenggi12_1,Haenggi14twc} for Aloha-based ad hoc networks. Next, we analyze the second-order statistics of the interference in MIMO HetNets.

% In Journal version
\section{Second-Order Statistics of HetNet Interference}\label{sec:int_stats}

The performance of diversity combining is fundamentally limited by the nature of the channel observed by the receiver. More specifically, variability of the reception quality on the one hand, and degree of its correlation across Rx antennas on the other, dictate how much can be gained by such techniques. Clearly, if the reception quality fluctuates considerably and independently across antennas, large gains can be expected. If, in contrast, the channel quality remains the same or does not vary significantly across Rx antennas, little to no gains can be expected. From \eqref{eq:mimo_gen1} it is evident that the resulting per-antenna reception quality is not only affected by the fading on the desired link but also by interference. Being a dynamic quantity, the latter contributes to the overall correlation structure and to variability of the reception quality with its own statistical properties. Compared to the influence from the desired channel, this contribution is yet relatively unexplored, particularly within the context of HetNets. This motivates to take a closer look at the second-order characteristics of the interference experienced at the typical user.

Let $\mathbf h_{i,n}=[\h_{i,n1},\ldots,\h_{i,nM_k}]$ be the $n\th$ row of $\mathbf H_i$. Then, the interference in slot $\tau$ (we drop the index $\tau$ in the following) at the $n\th$ Rx antenna, averaged over the code symbols $\mathbf c_i$ in one frame, is
\begin{IEEEeqnarray}{rCl}
   \mathsf{I}_{n}&=&\mathbb{E}_{\mathbf{c}_i}\left[\left(\sum\limits_{k=1}^{K}\sum\limits_{\x_i\in\Phi_k^{o}}\mathbf h_{i,n}\mathbf{c}_{i}\right)\left(\sum\limits_{k=1}^{K}\sum\limits_{\x_i\in\Phi_k^{o}}\mathbf h_{i,n}\mathbf{c}_{i}\right)^{H}\right]\IEEEnonumber\\
   &\overset{\text{(a)}}{=}&\sum\limits_{k=1}^{K}\sum\limits_{\x_i\in\Phi_k^{o}}\mathbf h_{i,n}\mathbb{E}\left[\mathbf c_{i}\mathbf c_i^{H}\right]\mathbf{h}_{i,n}^{H}\IEEEnonumber\\
   &\overset{\text{(b)}}{=}&\sum\limits_{k=1}^{K}\sum\limits_{\x_i\in\Phi_k^{o}}\frac{P_k}{S_k\|\x_i\|^{\alpha_k}}\mathbf h_{i,n}\text{diag}(\mathbf{v}_{i,\tau})\mathbf{h}_{i,n}^{H}\IEEEnonumber\\
   &=&\sum\limits_{k=1}^{K}\sum\limits_{\x_i\in\Phi_k^{o}}\frac{P_k}{S_k\|\x_i\|^{\alpha_k}}\|\mathbf h_{i,n}(S_k)\|_{F}^{2}=\sum\limits_{k=1}^{K}\mathsf{I}_{n,k},\IEEEeqnarraynumspace\label{eq:int_gen}
\end{IEEEeqnarray}
where (a) follows from the independence between the $\mathbf c_i$ across BSs and (b) follows from the correlation properties of the $\mathbf c_i$.

\subsection{Normalized Interference Variance}
Combining \eqref{eq:int_gen} and \eqref{eq:mimo_gen1} we can see that the dynamic part not belonging to the fading on the desired channel is $\mathsf{I}_n+\sigma^2$. Since this term is essentially a function of the long-term received useful power $P_\ell y^{-\alpha_\ell}$ through the cell association rule, it is reasonable to consider the normalized version $(\mathsf{I}_n+\sigma^2)/(P_\ell y^{-\alpha_\ell})\mathdef\I_n'$. This expression can be intuitively understood as the interference-to-average-signal ratio in branch $n$. Given that the serving BS is at distance $y$ and belongs to tier $\ell$, the corresponding conditional variance seen by the typical user is
\begin{IEEEeqnarray}{rCl}
   \var_{\ell,y}\left[\I_n'\right]&\overset{\text{(a)}}=&\frac{y^{2\alpha_\ell}}{P_\ell^2}\sum\limits_{k=1}^{K}\var_{\ell,y}\left[\mathsf{I}_{n,k}\right]\IEEEnonumber\\
   &\overset{\text{(b)}}{=}&2\pi y^{2\alpha_\ell}\sum\limits_{k=1}^{K}\hat P_k^2\lambda_k\frac{\mathbb{E}[\|\mathbf h_{n}(S_k)\|_{F}^{4}]}{S_k^2}\int_{d_k}^{\infty}\hspace{-.1cm}r^{-2\alpha_k+1}\,\mathrm dr\IEEEnonumber\\
   &\overset{\text{(c)}}{=}&\pi\sum\limits_{k=1}^{K}\frac{\lambda_k \hat P_k^{2/\alpha_k}}{\alpha_k-1}\left(1+\frac{1}{S_k}\right)y^{2/\hat\alpha_k},\IEEEeqnarraynumspace\label{eq:cond_var}
\end{IEEEeqnarray}
where (a) follows from the fact $\text{Var}[\mathsf{z}+c]=\text{Var}[\mathsf{z}]$ and from the independence of the $\mathsf{I}_{n,k}$ across tiers, (b) follows from Campbell's theorem\cite{stoyan95} and from the radius $d_k$ of the exclusion ball for the $k\th$ tier due to the cell association rule, and (c) follows from Lemma~\ref{lem:fading_dis}. The expression in \eqref{eq:cond_var} reflects the variance of the normalized interference for a given network geometry around the typical user. We thus still need to de-condition on $y$ and the associated tier $\ell$, which can be done using Lemma~\ref{lem:asso}. To obtain a more tractable expression that reveals the underlying trend, we shall assume equal path loss exponents and the same number of active Tx antennas across tiers, i.e., $\alpha_k\equiv\alpha$ and $S_k\equiv S$, next. The variance experienced at the typical user then becomes
\begin{IEEEeqnarray}{rCl}
   \var\left[\I_n'\right]
   &=&\sum\limits_{\ell=1}^{K}2\pi\lambda_{\ell}\int_{0}^{\infty}y\,\var_{\ell,y}\left[\I_n'\right]\IEEEnonumber\\
   &&\qquad\times\exp\left(-\pi y^2\sum\limits_{k=1}^{K}\lambda_k \hat{P}_k^{2/\alpha}\right)\mathrm dy=\frac{1+\frac{1}{S}}{\alpha-1}.\IEEEeqnarraynumspace\label{eq:int_var}
\end{IEEEeqnarray}

The following observations can be made from \eqref{eq:int_var}:
\begin{itemize}
    \item Interestingly, the variance of the normalized interference neither depends on the number of tiers nor on their parameters $P_k$ and $\lambda_k$ when $\alpha_k\equiv\alpha$ and $S_k\equiv S$. This is consistent with\cite{jo12,dhillon12}, where the independence between the $\sinr$ distribution and the number/parameters of tiers was shown for equal $\alpha_k$ and absence of Rx noise.
    \item In line with our intuition, the interference variance increases when $\alpha$ becomes smaller as the interference contribution from far-off BSs carries more weight. Conversely, for large $\alpha$ the interference is dominated by a few close-by BSs, which reduces the interference variance. For typical path loss exponents around $\alpha=3.7$\cite{3gpp_tr_36814}, the variance is $0.74$ when $S=1$. 
    \item With the same path loss law $\|\cdot\|^{-\alpha}$, the interference variance in Aloha-based ad hoc networks diverges\cite{HaenggiBook}, as interfering transmitters can be arbitrarily close to the receiver in this case. Although this result is due to the singularity of the path loss law and has no physical relevance, it follows that the interference variance in HetNets tends to be smaller than in Aloha-based ad hoc networks. This, in turn, suggests that IA diversity combining will generally perform lower in HetNets. We shall discuss this further in Section~\ref{sec:mrc_simo}.
    \item The interference variance decays with the number of active Tx antennas $S$. This is because adding more Tx antennas while reducing the per-antenna Tx power by $S$ smoothes out the channel fluctuations. This effect is also referred to as {\it channel-hardening}\cite{hochwald04}. Hence, for large $S$ we expect IA schemes to perform similar to IB schemes due to less interference variability. In the limit $S\to\infty$, the interference variance becomes $\tfrac{1}{\alpha-1}$.
 \end{itemize}
 
\subsection{Normalized Interference Correlation across Antennas}
 
To study the interference correlation across Rx antennas, we first need the covariance of the normalized interference, conditioned on the serving tier $\ell$ and BS distance $y$. The covariance is
\begin{IEEEeqnarray}{rCl}
    &&\cov_{\ell,y}\left[\I_u',\I_v'\right]\IEEEnonumber\\
    &&\qquad\overset{\text{(a)}}{=}\frac{y^{2\alpha_\ell}}{P_\ell^2}\sum\limits_{k=1}^{K}\cov_{\ell,y}\left[\mathsf{I}_{u,k},\mathsf{I}_{v,k}\right]\IEEEnonumber\\
    &&\qquad\overset{\text{(b)}}{=}2\pi y^{2\alpha_\ell}\sum\limits_{k=1}^{K}\hat P_k^2\lambda_k\frac{\mathbb{E}[\|\mathbf h_u(S_k)\|_{F}^2]^2}{S_k^2}\int_{d_k}^{\infty}r^{-2\alpha_k+1}\,\mathrm dr\IEEEnonumber\\
    &&\qquad\overset{\text{(c)}}{=}\pi\sum\limits_{k=1}^{K}\frac{\lambda_k \hat P_k^{2/\alpha_k}}{\alpha_k-1}y^{2/\hat\alpha_k},\IEEEeqnarraynumspace\label{eq:cond_cov}
\end{IEEEeqnarray}
 where (a) follows from $\cov[\mathsf{z}_{1}+c,\mathsf{z}_{2}+c]=\cov[\mathsf{z}_1,\mathsf{z}_2]$ and from the independence of the $\mathsf{I}_{n,k}$ across $k$, (b) follows from Campbell's theorem\cite{stoyan95} and from $\mathsf{h}_n$ being independent across $n$, and (c) follows from Lemma~\ref{lem:fading_dis}. With \eqref{eq:cond_var} and \eqref{eq:cond_cov}, the conditional correlation coefficient becomes
 \begin{IEEEeqnarray}{rCl}
    \rho_{\ell,y}&=&\frac{\cov_{\ell,y}\left[\I_u',\I_v'\right]}{ \var_{\ell,y}\left[\I_n'\right]}=\frac{\cov_{\ell,y}\left[\I_u,\I_v\right]}{\var_{\ell,y}\left[\I_n\right]}\overset{S_k\equiv S}{=}\frac{S}{1+S}.\label{eq:cond_cor}\IEEEeqnarraynumspace
\end{IEEEeqnarray}

The following observations can be made from \eqref{eq:cond_cor}:
  \begin{itemize}
 	\item The correlation coefficient $\rho_{\ell,y}$ has the same form as the {\it temporal} correlation coefficient in Aloha-based ad hoc networks derived in \cite[Lem.~5.13]{HaenggiBook}. Temporal correlation with fixed set of active interferers and spatial correlation across Rx antennas are mathematically the same, since in both cases fading varies while the interferer locations remain fixed.
 	
 	\item As expected, adding more Tx antennas increases the interference correlation across Rx antennas, since the channel fluctuations then undergo an averaging effect. For $S=1$ we have $\rho_{\ell,y}=1/2$. In the limit $S\to\infty$, the interference correlation becomes maximal, i.e., $\rho_{\ell,y}=1$. Similar to the above comment on interference variance for large $S$, IA schemes are expected to have the same performance as IB schemes in this regime.
 
 	\item Interestingly, $\rho_{\ell,y}$ is independent from the tier with which the typical user associates and also from the distance to the serving BS when the number of active Tx antennas is equal across tiers ($S_k\equiv S$). In this case, $\rho_{\ell,y}$ is unaffected by $K$, $P_k$, $\lambda_k$, and, contrary to the interference variance, also independent from the path loss exponents $\alpha_k$. %his result is somewhat remarkable since, as will be shown later in Section~\ref{sec:cp_analysis}, the effect of correlation on the performance  is not.
\end{itemize}

 \begin{remark}[Feasibility of Interference Estimation]\label{rem:int_pwr_est}
	When the set of active antennas of interfering BSs changes in every slot $\tau$, $\mathsf{I}_{n}$ varies unpredictably between every slot of the codeword. This is the case when $S_k<M_k$. Such rapid variations over $\tau$ are imperceptible to CSI estimation since the latter is usually designed to track channel-fading variations, which happen on a larger time scale. However, when full-rate OSTBCs are used ($r_k=1$ for all $k$), $\mathsf{I}_{n}$ is identical across $\tau$, since all $S_k=M_k$ Tx antennas are always active. In that case, the receiver can obtain knowledge of $\mathsf{I}_{n}+\sigma^2$ with acceptable complexity, e.g., after decoding and removing the pilot symbols sent by the serving BS \cite{gosh10} or by using techniques from \cite{benedict67,pauluzzi00}.
 \end{remark}

\section{Coverage Probability Analysis}\label{sec:cp_analysis}
We now study the downlink performance at the typical user for both IB-MRC and IA-MRC. 
%As explained in Remark~\ref{rem:int_pwr_est}, IA-MRC is practical only for full-rate OSTBCs ($r_k=1$ for all $k$). 
A common way for studying the performance of diversity-combining techniques is to analyze the post-combiner signal-to-interference-plus-noise ratio ($\sinr$). The specific form of the $\sinr$ depends on the considered scheme and will be developed in \ref{sec:int_blind} and \ref{sec:int_aware}.

\newcounter{mycounter1}
\newcounter{mycounter2}
\begin{figure*}[!t]
\normalsize
\setcounter{equation}{17}
\begin{IEEEeqnarray}{rCl}	   \cpib&=&2\pi\sum\limits_{\ell=1}^{K}\sum\limits_{m=0}^{NM_\ell-1}\hspace{-.1cm}\frac{(-1)^m \lambda_{\ell}}{m!}\hspace{-.1cm}\int_{0}^{\infty}\hspace{-.15cm}y\,\frac{\mathrm d^m}{\mathrm ds^m}\left[\exp\left(-\frac{sS_\ell T}{\snr_\ell(y)}-\pi\sum\limits_{k=1}^{K}\lambda_k \hat P_k^{2/\alpha_k}y^{2/\hat \alpha_k}\hypergeom{-\frac{2}{\alpha_k}}{S_k}{1-\frac{2}{\alpha_k}}{-\frac{s T}{\hat S_k}}\right)\right]_{s=1}\hspace{-.4cm}\mathrm dy,\IEEEeqnarraynumspace\label{eq:cp_ib}
\end{IEEEeqnarray}
\setcounter{mycounter1}{18}
\hrulefill
\normalsize
\setcounter{equation}{27}
\begin{IEEEeqnarray}{rCl}	   \cpia&=&2\pi\sum\limits_{\ell=1}^{K}\sum\limits_{m=0}^{M_\ell-1}\frac{(-1)^{m+M_\ell}\lambda_{\ell}}{m!\,\Gamma(M_\ell)}\int_{0}^{\infty}\int_{0}^{\infty}yz^{-1}\frac{\mathrm d^{m+M_\ell}}{\mathrm ds^{m}\,\mathrm dt^{M_\ell}}\bigg[ \exp\left(-\frac{M_\ell}{\snr_\ell(y)}\left(s\,(T-z)^{+}+tz\right)\right)\IEEEnonumber\\
&&\hspace{4.5cm}\times\exp\left(-\pi\sum\limits_{k=1}^{K}\lambda_k \hat P_k^{2/\alpha_k}y^{2/\hat \alpha_k} \left[1+\Psi\left(\frac{s\,(T-z)^+}{\hat M_k},\frac{tz}{\hat M_k},M_k,\alpha_k\right)\right]\right)\bigg]_{\substack{s=1\\t=1}}\hspace{-.0cm}\mathrm dy\,\mathrm dz,\IEEEeqnarraynumspace\label{eq:cp_ia}
\end{IEEEeqnarray}
\hrulefill
\setcounter{mycounter2}{28}
\setcounter{equation}{11}
\vspace*{4pt}
\end{figure*}

\begin{definition}[Coverage Probability $\cp$]\label{def:cp}
 The coverage probability is defined as
\begin{IEEEeqnarray}{rCl}
	\cp\mathdef\mathbb{P}\left(\sinr\geq T\right)\label{eq:cp_general}
\end{IEEEeqnarray}
for a coding and modulation specific $\sinr$ threshold $T>0$.
\end{definition}
The $\cp$ can be interpreted as the complementary CDF of $\sinr$ at the typical user, or alternatively as the average fraction of users in the HetNet covered by an $\sinr$ no less than $T$.

\subsection{MIMO Diversity with interference-blind MRC}\label{sec:int_blind}

A useful feature of OSTBCs is that the MIMO channel \eqref{eq:mimo_gen2} can be reduced to parallel SISO channels\cite{paulray03}. At the typical user, knowing $\mathbf{H}_{o}$, this is achieved by the linear combination
\begin{IEEEeqnarray}{rCl}
   \sum_{n=1}^{N}\sum_{m=1}^{M_\ell}\h_{o,nm}^{\ast}\mathbf{A}_{nm}^{H}\mathbf{\bar r}+\h_{o,nm}\mathbf{B}_{nm}^{T}\mathbf{\bar r}^{\ast}\label{eq:lin_com_ib}
\end{IEEEeqnarray}
where $\mathbf{A}_{nm}$ and $\mathbf{B}_{nm}$ are the $NL_\ell\times S_\ell$ dispersion matrices describing the OSTBC employed in the serving tier, see\cite{tarokh99,shang08} for further details. The resulting {\it equivalent channel model} allows treating the detection of each of the $S_\ell$ information symbols of the current codeword separately. The corresponding $\sinr$ at the symbol decoder can then be expressed as
\begin{IEEEeqnarray}{rCl}
   \sinr_{\ell}(y)&=&\frac{\frac{P_{\ell}}{S_\ell y^{\alpha_{\ell}}}\|\mathbf{H}_o\|_{F}^2}{\sum_{k=1}^{K}\sum_{\x_i\in\Phi^{o}_k}\mathsf{I}_{i,\text{eqv}}+\sigma^2},\label{eq:sinr_ib}\IEEEeqnarraynumspace
\end{IEEEeqnarray}
where $\mathsf{I}_{i,\text{eqv}}$ is the interference from the $i\th$ BS in the equivalent channel model. $\mathsf{I}_{i,\text{eqv}}$ is statistically the same for all $S_\ell$ symbols. Thus, focusing on an arbitrary symbol, i.e., considering a single arbitrary column of $\mathbf{A}_{nm}$, $\mathbf{B}_{nm}$, say $\mathbf{a}_{nm}$, $\mathbf{b}_{nm}$, the interference $\mathsf{I}_{i,\text{eqv}}$ is
\begin{IEEEeqnarray}{rCl}
   \mathsf{I}_{i,\text{eqv}}&=&\var_{\mathbf c_i}\hspace{-.1cm}\left[\sum\limits_{n=1}^{N}\sum\limits_{m=1}^{M_\ell}\frac{\h_{o,nm}^{\ast}}{\|\mathbf H_o\|_F}\mathbf{a}_{nm}^{H}\mathbf{\bar i}_i+\frac{\h_{o,nm}}{\|\mathbf H_o\|_F}\mathbf{b}_{nm}^{T}\mathbf{\bar i}_i^{\ast}\right].\IEEEeqnarraynumspace\label{eq:int_blind_mrc}
\end{IEEEeqnarray}

Note that the Rx noise statistics remain unaffected by the linear combination in \eqref{eq:lin_com_ib}\cite{paulray03, shang08}. However, the distribution of $\mathsf{I}_{i,\text{eqv}}$ is more complicated, particularly due to its dependence on $\mathbf H_o$. This was already observed in\cite{hunter08} for a similar MIMO network model, where the authors also showed that ignoring this dependence and assuming $\mathsf{I}_{i,\text{eqv}}$ to be Gamma distributed yields a valid approximation. We thus follow the same approach and assume $\mathsf{I}_{i,\text{eqv}}\simeq \tfrac{P_k}{S_k\|\x_i\|^{\alpha_k}}\|\mathbf{H}_i(S_k)\|_F^2$ with $\mathsf{I}_{i,\text{eqv}}$ being independent from $\mathbf H_o$, which can be viewed as effectively ignoring the effect of the MIMO processing on interference. The following two facts support this approximation:
\begin{itemize}
   \item It can be shown that the approximation is moment matching irrespective of the realization of $\mathbf H_o$, i.e., $\mathbb{E}_{\mathbf{H}_{i}}[\mathsf{I}_{i,\text{eqv}}]=\tfrac{P_k}{S_k\|\x_i\|^{\alpha_k}}\mathbb{E}_{\mathbf{H}_{i}}[\|\mathbf{H}_i(S_k)\|_F^2]=\tfrac{P_k}{\|\x_i\|^{\alpha_k}}$ in \eqref{eq:int_blind_mrc}.
   \item Whenever $M_k=1$, it follows from\cite{shah00} that the above approximation becomes exact. In this case $\mathsf I_{i,\text{eqv}}$ is also truly independent from $\mathbf H_o$.
\end{itemize}%\vspace{-8pt}

\begin{lemma}[Interference Laplace Transform]\label{lem:int_lap}
   Consider the interference field $\mathsf{I}=\sum_{k=1}^{K}\sum_{\x_i\in\Phi^{o}_k}$ $\tfrac{P_k}{S_k\|\x_i\|^{\alpha_k}}\|\mathbf{H}_i(S_k)\|_F^2$. Its Laplace transform is given by
   \begin{IEEEeqnarray}{rCl}
      &&\lap_{\mathsf{I}}(s)=e^{-\pi\sum\limits_{k=1}^{K}\lambda_kd_k^2\left(\hypergeomexp{-\tfrac{2}{\alpha_k}}{S_k}{1-\tfrac{2}{\alpha_k}}{-\tfrac{sP_k}{S_kd_k^{\alpha_k}}}-1\right)}.\label{eq:int_lap}
      \IEEEeqnarraynumspace
   \end{IEEEeqnarray}
\end{lemma}%\vspace{-4pt}
\begin{IEEEproof}
   We write%\vspace{-4pt}
   \begin{IEEEeqnarray}{rCl}
      &&\mathbb{E}\left[\exp\left(-s\sum_{k=1}^{K}\sum_{\x_i\in\Phi^{o}_k}\frac{P_k}{S_k\|\x_i\|^{\alpha_k}}\|\mathbf{H}_i(S_k)\|_F^2\right)\right]\IEEEnonumber\\
      &&\quad\overset{\text{(a)}}{=}\prod\limits_{k=1}^{K}\mathbb{E}\left[\prod\limits_{\x_i\in\Phi_k^o}\lap_{\|\mathbf{H}_i(S_k)\|_F^2}\left(\frac{sP_k}{S_k\|\x_i\|^{\alpha_k}}\right)\right]\IEEEnonumber\\
      &&\quad\overset{\text{(b)}}{=}\prod\limits_{k=1}^{K}\mathbb{E}\left[\prod\limits_{\x_i\in\Phi_k^o}\left(1+\frac{sP_k}{S_k\|\x_i\|^{\alpha_k}}\right)^{-S_k}\right]\IEEEnonumber\\
      &&\quad\overset{\text{(c)}}{=}\exp\left\{-\pi\sum\limits_{k=1}^{K}\lambda_k\int_{d_k}^{\infty}\hspace{-.1cm}2r\left(1-\left(1+\tfrac{sP_k}{S_kr^{\alpha_k}}\right)^{-S_k}\right)\mathrm dr\right\},\IEEEnonumber\\
   \end{IEEEeqnarray}
	where (a) follows from the independence of the $\Phi_k^o$ across $k$ and from the independence of the $\|\mathbf{H}_i(S_k)\|_F^2$ across $i$, (b) follows from the Laplace transform of the Erlang distributed $\|\mathbf{H}_i(S_k)\|_F^2$, and (c) follows from the probability generating functional (PGFL) for PPPs, see \cite{stoyan95}.% Solving the integral yields the result.
\end{IEEEproof}

We now have the tools required to characterize the coverage probability for IB-MRC.% This task is addressed in the following theorem.

\begin{theorem}[$\cp$ for IB-MRC]\label{thm:cp_ib}
   The coverage probability for IB-MRC in the described setting is given by \eqref{eq:cp_ib} at the top of the next page, 
%    \begin{IEEEeqnarray}{rCl} \cpib&=&2\pi\sum\limits_{\ell=1}^{K}\sum\limits_{m=0}^{NM_\ell-1}\frac{(-1)^m \lambda_{\ell}}{m!}\int_{0}^{\infty}y\, \frac{\mathrm d^m}{\mathrm ds^m}\left[\exp\left(-\frac{sS_\ell T}{\snr_\ell(y)}\right.\right.\IEEEnonumber\\
%    &&\left.\left.\hspace{4cm}-\pi\sum\limits_{k=1}^{K}\lambda_k \hat P_k^{2/\alpha_k}y^{2/\hat \alpha_k}\hypergeom{-\frac{2}{\alpha_k}}{S_k}{1-\frac{2}{\alpha_k}}{-\frac{s T}{\hat S_k}}\right)\right]_{s=1}\hspace{-.5cm}\mathrm dy,\IEEEeqnarraynumspace\label{eq:cp_ib}
% \end{IEEEeqnarray}
  where $\snr_{\ell}(y)\mathdef P_\ell\, y^{-\alpha_\ell}/\sigma^2$ and $\hat S_k\mathdef S_k/S_\ell$.
\end{theorem}
\begin{IEEEproof}
   See Appendix~\ref{ap:cp_ib}.
\end{IEEEproof}
\setcounter{equation}{\value{mycounter1}}

The Gaussian hypergeometric function $\hypergeom{\cdot}{\cdot}{\cdot}{\cdot}$ can be expressed through elementary functions for certain $\alpha_k$\cite{olver10}. For instance, $\hypergeomexp{1}{-\tfrac{1}{2}}{\tfrac{1}{2}}{-u}=1+\sqrt{u}\arctan\sqrt{u}$ when $\alpha_k=4$. For general $\alpha_k>2$, \eqref{eq:cp_ib} can be evaluated using numerical software programs, see Remark~\ref{rem:numerical}.

The derivative $\mathrm d^m/\mathrm ds^m$ in \eqref{eq:cp_ib} can be calculated using Fa\`{a} di Bruno's formula for higher-order derivatives of composite functions\cite{olver10}, i.e., with an inner and outer function. While the outer function of the integrand is simple due to the $\exp$-term, the inner function, i.e., ${}_2F_1(-2/\alpha_k,S_k,1-2/\alpha_k;-sT/\hat S_k)$, is more involved. With\cite{olver10}, its derivative is obtained as
\begin{IEEEeqnarray}{rCl}
   &&\frac{\mathrm d^m}{\mathrm ds^m}\left[\hypergeom{-\tfrac{2}{\alpha_k}}{S_k}{1-\tfrac{2}{\alpha_k}}{-\tfrac{sT}{\hat S_k}}\right]_{s=1}\IEEEnonumber\\
   &&\quad=\left(-\tfrac{T}{\hat S_k}\right)^m\frac{-2/\alpha_k\,\Gamma(S_k+m)}{(m-2/\alpha_k)\,\Gamma(S_k)}\IEEEnonumber\\
   &&\qquad\times\hypergeom{-\tfrac{2}{\alpha_k}+m}{S_k+m}{1-\tfrac{2}{\alpha_k}+m}{-\tfrac{T}{\hat S_k}}.\IEEEeqnarraynumspace\label{eq:diff}
\end{IEEEeqnarray}

In dense deployments the performance is typically limited by interference rather than noise\cite{goldsmith05}, which yields $\sigma^2=0\Leftrightarrow1/\snr_\ell(y)=0$ for all $\ell,y$. In addition, the path loss exponent does not vary significantly across tiers in practice with typical values around $\alpha_k\approx 3.7$ \cite{3gpp_tr_36814}. If, in addition, all tiers have the same number of Tx antennas, the following corollary applies.

\begin{corollary}[Special Case]\label{col:cp_ib_col1}
   In the absence of Rx noise ($\sigma^2=0$) and with equal path loss exponents ($\alpha_k\equiv\alpha$) and number of Tx antennas ($M_k\equiv M$, $S_k\equiv S$), $\cpib$ reduces to
   \begin{IEEEeqnarray}{rCl}
      \cpib&=&\sum\limits_{m=0}^{NM-1}\frac{(-1)^m}{m!}\frac{\mathrm d^m}{\mathrm ds^m}\Big[\frac{1}{\hypergeom{-\frac{2}{\alpha}}{S}{1-\frac{2}{\alpha}}{-sT}}\Big]_{s=1}\hspace{-.05cm}.\IEEEeqnarraynumspace\label{eq:cp_ib_col1}
   \end{IEEEeqnarray}
\end{corollary}

The coverage probability in \eqref{eq:cp_ib_col1} neither depends on the BS densities $\lambda_k$ and powers $P_k$, nor on the number of tiers $K$, which is consistent with the literature, see for instance\cite{dhillon12}. Note that the first term $m=0$ in \eqref{eq:cp_ib_col1} corresponds to the coverage probability for the SISO case\cite{jo12}.%\vspace{-6pt}

\subsection{MIMO Diversity with interference-aware MRC}\label{sec:int_aware}

We now assume $M_k\leq 2$ for all tiers, which ensures that the receiver can estimate the interference-plus-noise power at each Rx antenna with acceptable complexity once within the current block/frame, see Remark~\ref{rem:int_pwr_est} for more details. Note that, in theory, $M_k> 2$ is also possible though not practical as the estimation would then have to be performed in each slot $\tau$ in the presence of the desired code symbols $\mathbf{c}_{o,\tau}$ to be decoded. When $M_k\leq 2$, we have $S_k=M_k$ for all $k$, meaning that either Alamouti space-time coding ($M_k=2$) or no space-time coding ($M_k=1$) is used in tier $k$. In both cases, the total interference then remains constant during the entire codeword. Its value at the $n\th$ Rx antenna is given by \eqref{eq:int_gen}. We assume that the receiver knows the current per-antenna interference-plus-noise power $\mathsf{I}_n+\sigma^2$ at each Rx antenna, in addition to $\mathbf{H}_o$. Interference is still treated as white noise. In IA-MRC, the phase-corrected and channel-weighted received signals are normalized by $\mathsf{I}_n+\sigma^2$ at each Rx antenna, thereby following the original MRC approach from \cite{brennan59}. The receiver hence performs the linear combination
\begin{IEEEeqnarray}{rCl}
   \sum\limits_{n=1}^{N}\sum\limits_{m=1}^{M_\ell}\frac{\h_{o,nm}^\ast}{\I_n+\sigma^2}\mathbf{A}^{H}_{nm}\mathbf{\bar r}+\frac{\h_{o,nm}}{\I_n+\sigma^2}\mathbf{B}^{T}_{nm}\mathbf{\bar r}^\ast,\label{eq:lin_com_ia}
\end{IEEEeqnarray}
yielding the equivalent channel model for IA-MRC. %For instance, when $M_k=1$ ($L_k=1$, no space-time coding) for all $K$ tiers, \eqref{eq:lin_com_ia} reduces to
% \begin{IEEEeqnarray}{rCl}
%    \mathsf{c}_{o}\sum\limits_{n=1}^{N}\frac{|h_{o,n}|^2}{I_{n}+\sigma^2}+\sum\limits_{k=1}^{K}\sum_{\x_i\in\Phi_k^o}\mathsf{c}_i\sum\limits_{n=1}^{N}\frac{h_{o,n}^{\ast}\h_{i,n}}{I_{n}+\sigma^2}+\sum\limits_{n=1}^{N}\frac{h_{o,n}^{\ast}\mathsf{n}_{n}}{I_{n}+\sigma^2},\IEEEeqnarraynumspace
% \end{IEEEeqnarray}
%where $\mathsf{c}_{i}$ represents the symbol transmitted by the $i\th$ BS. A similar expression can be found for $M_\ell=2$ and general $M_k\leq 2$ by using the dispersion matrices $\mathbf{A}_{nm}$, $\mathbf{B}_{nm}$ for Alamouti, which are available for instance in\cite{shah00}. 
%Similar to Section~\ref{sec:int_blind}, 
We focus again on an arbitrary symbol and consider an arbitrary column $\mathbf{a}_{nm}$, $\mathbf{b}_{nm}$ of $\mathbf{A}_{nm}$, $\mathbf{B}_{nm}$. The $\sinr$ can then be given as
\begin{IEEEeqnarray}{rCl}
   \sinr_{\ell}(y)&=&\frac{\frac{P_\ell}{M_\ell\, y^{\alpha_\ell}} \left(\sum_{n=1}^{N}\frac{\|\mathbf{h}_{o,n}\|_{F}^2}{\mathsf{I}_n+\sigma^2}\right)^2}{\sum_{k=1}^{K}\sum_{\x_i\in\Phi^{o}_k}\mathsf{I}_{i,\text{eqv}}+\sum_{n=1}^{N}\frac{\|\mathbf{h}_{o,n}\|_F^2\sigma^2}{(\mathsf{I}_n+\sigma^2)^2}},\label{eq:sinr_ia}\IEEEeqnarraynumspace
\end{IEEEeqnarray}
where now $\mathsf{I}_{i,\text{eqv}}$ is
\begin{IEEEeqnarray}{rCl}
   \mathsf{I}_{i,\text{eqv}}
   &=&\var_{\mathbf{c}_i}\hspace{-.1cm}\left[\sum\limits_{n=1}^{N}\sum\limits_{m=1}^{M_\ell}\frac{\h_{o,nm}^\ast}{\mathsf{I}_n+\sigma^2}\mathbf{a}^{H}_{nm}\mathbf{\bar i}_i+\frac{\h_{o,nm}}{\mathsf{I}_n+\sigma^2}\mathbf{b}^{T}_{nm}\mathbf{\bar i}_i^\ast\right].\IEEEeqnarraynumspace\label{eq:int_ia}
\end{IEEEeqnarray}

By \eqref{eq:mimo_gen2}, $\mathbb{E}_{\mathbf{c}_i}[\mathbf{\bar i}_i \mathbf{\bar i}_i^{H}]=\frac{P_k}{M_k\|\x_i\|^{\alpha_k}}\left(\|\mathbf{h}_{i,n}\|_F^2\mathbf{I}_{NL_\ell}+\mathbf{R}_{i}\right)$, where $\mathbf{R}_{i}=\text{diag}(\mathbf{\tilde H}_i,\ldots,\mathbf{\tilde H}_i)$ is a block diagonal matrix with the $N\times N$ square matrix $\mathbf{\tilde H}_i$ on the diagonal. The entries of $\mathbf{\tilde H}_i$ are
\begin{subnumcases}{\label{eq:h_tilde_mat} (\mathbf{\tilde H}_i)_{pq}=} 
                               \mathbf{h}_{i,p}\mathbf{h}_{i,q}^{H},\quad p\neq q\\
                               0, \quad p=q.
\end{subnumcases}
Note that, in contrast to $\mathbf{I}_{NL_\ell}$, $\mathbf{R}_{i}$ has non-zero off-diagonal matrix entries. Invoking the orthogonality properties of $\mathbf{A}_{nm},\mathbf{B}_{nm}$, i.e., $\mathbf{A}_{nm}^{H}\mathbf{A}_{uv}+\mathbf{B}_{uv}^{T}\mathbf{B}_{nm}^{\ast}=\delta_{nu}\delta_{mv}\mathbf{I}_{NL_\ell}$ and $\mathbf{A}_{nm}^{H}\mathbf{B}_{uv}+\mathbf{B}_{uv}^{T}\mathbf{A}_{nm}^{\ast}=\mathbf{0}_{L_\ell}$, where $\delta_{ij}=1$ if $i=j$ and zero otherwise, and exploiting the mathematical structure of $\mathbb{E}_{\mathbf{c}_i}[\mathbf{\bar i}_i \mathbf{\bar i}_i^{H}]$, \eqref{eq:int_ia} can be computed following the same approach as in \cite[Sec.~2.2.3]{shang08} as 
\begin{IEEEeqnarray}{rCl}
   \mathsf{I}_{i,\text{eqv}}&=&\frac{P_k}{M_k\|\x_i\|^{\alpha_k}}\hspace{-2pt}\sum\limits_{n=1}^{N}\hspace{-1.5pt}\|\mathbf{h}_{o,n}\|_F^2\frac{\|\mathbf{h}_{i,n}\|_F^2}{(\mathsf{I}_n+\sigma^2)^2}+\frac{\mathsf{Z}_{i,n}}{(\mathsf{I}_n+\sigma^2)^2},\IEEEeqnarraynumspace\label{eq:int_ia2}
\end{IEEEeqnarray}
where $\mathsf{Z}_{i,n}$ describes the part resulting from $\mathbf{R}_{i}$ having non-zero off-diagonal matrix entries. From \eqref{eq:h_tilde_mat}, it can be inferred that $\mathsf{Z}_{i,n}$ depends on the channel phases of $\mathbf{H}_i$. Since $\h_{i,nm}\sim\mathcal{CN}(0,1)$, it can be shown that $\mathbb{E}_{\angle\mathbf{H}_i}[\mathbf{R}_{i}]=0$, which implies $\mathbb{E}_{\angle\mathbf{H}_i}[\mathsf{Z}_{i,n}]=0$ irrespective of $\mathbf{H}_{o}$, indicating that the effect of $\mathsf{Z}_{i,n}$ vanishes ``in the long run''. To obtain a more tractable expression, we hence ignore $\mathsf{Z}_{i,n}$. With this simplification and after some algebraic manipulations, \eqref{eq:sinr_ia} then becomes
\begin{IEEEeqnarray}{rCl}
   \sinr_{\ell}(y)&=&\frac{P_\ell}{M_\ell\,y^{\alpha_\ell}} \sum_{n=1}^{N}\frac{\|\mathbf{h}_{o,n}\|_{F}^2}{\mathsf{I}_n+\sigma^2}.\label{eq:sinr_ia_simple}\IEEEeqnarraynumspace
\end{IEEEeqnarray}

\begin{remark}[$\sinr$-Approximation]\label{rem:sinr_approx} It follows by Jensen's inequality\cite{feller71} that
\begin{IEEEeqnarray}{rCl}
   \mathbb{E}_{\angle\mathbf{H}_1,\angle\mathbf{H}_2,\ldots}\left[\sinr_{\ell}(y)\right]&\geq&\frac{P_\ell}{M_\ell\,y^{\alpha_\ell}} \sum_{n=1}^{N}\frac{\|\mathbf{h}_{o,n}\|_{F}^2}{\mathsf{I}_n+\sigma^2},\IEEEeqnarraynumspace
\end{IEEEeqnarray}
where $\sinr_{\ell}(y)$ on the left-hand side is the exact $\sinr$ from \eqref{eq:sinr_ia}. Hence, the simplified $\sinr$ from \eqref{eq:sinr_ia_simple} is a lower bound on the phase-averaged exact $\sinr$. The error when \eqref{eq:sinr_ia_simple} is used to approximate the exact $\sinr$ is barely noticeable, as confirmed by simulations in Section~\ref{sec:design}.
\end{remark}

Although the $\mathbf h_{i,n}$ in \eqref{eq:sinr_ia_simple} are mutually independent, the $\mathsf{I}_{n}$ are correlated across Rx antennas due to the common locations of interfering BSs. More specifically, the expression in \eqref{eq:sinr_ia_simple} is a sum of correlated random variables exhibiting a complicated correlation structure. This renders the computation of the coverage probability for IA-MRC for general $N$ challenging. In practical systems, however, the number of antennas mounted on mobile devices is limited due to space/complexity limitations, thereby often not exceeding $N=2$. This case is addressed next.

\begin{theorem}[$\cp$ for IA-MRC]\label{thm:cp_ia}
	The coverage probability for IA-MRC in the described setting is given by \eqref{eq:cp_ia} at the top of the last page,
% 	\begin{IEEEeqnarray}{rCl}	  
% 	\cpia&=&2\pi\sum\limits_{\ell=1}^{K}\sum\limits_{m=0}^{M_\ell-1}\frac{(-1)^{m+M_\ell}\lambda_{\ell}}{m!\,\Gamma(M_\ell)}\int_{0}^{\infty}\int_{0}^{\infty}yz^{-1}\frac{\mathrm d^{m+M_\ell}}{\mathrm ds^{m}\,\mathrm dt^{M_\ell}}\bigg[ \exp\left(-\frac{M_\ell}{\snr_\ell(y)}\left(s\,(T-z)^{+}+tz\right)\right)\IEEEnonumber\\
% &&\hspace{.7cm}\times\exp\left(-\pi\sum\limits_{k=1}^{K}\lambda_k \hat P_k^{2/\alpha_k}y^{2/\hat \alpha_k} \left[1+\Psi\left(\frac{s\,(T-z)^+}{\hat M_k},\frac{tz}{\hat M_k},M_k,\alpha_k\right)\right]\right)\bigg]_{\substack{s=1\\t=1}}\hspace{-.0cm}\mathrm dy\,\mathrm dz,\label{eq:general_cp_1a}\IEEEeqnarraynumspace\label{eq:cp_ia}
% \end{IEEEeqnarray}
	where $1\leq M_k\leq 2,N=2$, $\hat M_k\mathdef M_k/M_\ell$, and $\Psi(\cdot,\cdot,\cdot,\cdot)$ is given by \eqref{eq:cp_proof_ia_step6} in Appendix~\ref{ap:cp_ia}.
\end{theorem}
\begin{IEEEproof}
See Appendix~\ref{ap:cp_ia}.
\end{IEEEproof}
\setcounter{equation}{\value{mycounter2}}

The function $\Psi(\cdot,\cdot,\cdot,\cdot)$ in \eqref{eq:cp_ia} can be given in terms of Gaussian hypergeometric functions $\hypergeom{\cdot}{\cdot}{\cdot}{\cdot}$, which can be further simplified in some cases, see comment after Theorem~\ref{thm:cp_ib} in Section~\ref{sec:int_blind}. %We shall exploit this fact in Section~\ref{sec:mrc_simo}.
%Compared to $\cpib$ in \eqref{eq:cp_ib}, $\cpia$ is more involved due to the mathematical form of \eqref{eq:sinr_ia_simple}, which translates into the convolution-type integral over $z$. %Nevertheless, the expression in \eqref{eq:cp_ia} can be evaluated with acceptable complexity using semi-analytical tools, see for instance\cite{tandhi14_2}. Besides, \eqref{eq:cp_ia} 
Theorem~\ref{thm:cp_ia} covers the general case and the expression can be further simplified as shown next.

 % Comment on relationship between IA and IB MRC here (journal version)

\begin{corollary}[Special Case]\label{col:cp_ia_col1}
   In the absence of Rx noise ($\sigma^2=0$), and with equal path loss exponents ($\alpha_k\equiv\alpha$) and number of Tx antennas ($M_k\equiv M\leq 2$), $\cpia$ from Theorem~\ref{thm:cp_ia} reduces to
   \begin{IEEEeqnarray}{rCl}
      \cpia&=&\sum\limits_{m=0}^{M-1}\frac{(-1)^{m+M}}{m!\,\Gamma(M)}\int_{0}^{\infty}z^{-1}\IEEEnonumber\\
      &&\times\frac{\mathrm d^{m+M}}{\mathrm ds^{m}\mathrm dt^{M}}\hspace{-.1cm}\left[\frac{1}{1+\Psi\left(s\,(T-z)^+,tz,M,\alpha\right)}\right]_{\substack{s=1\\t=1}}\mathrm dz.\IEEEeqnarraynumspace\label{eq:cp_ia_col1}
   \end{IEEEeqnarray}
\end{corollary}

\begin{remark}[Numerical Evaluation]\label{rem:numerical} The expressions in Theorem~\ref{thm:cp_ib}, Theorem~\ref{thm:cp_ia}, and in the related corollaries, can be evaluated using numerical tools. One possible approach is to first move the higher-order derivatives outside the integral(s), which is allowed by Leibniz's integral rule \cite{olver10}, to compute the integral(s) for specific $s$ (and $t$) using built-in algorithms available in numerical software programs. Thereby, $s$ (and $t$) are chosen as the Chebyshev nodes of the Chebyshev interpolation method \cite{press07}, which allows numerically computing the higher-order derivatives with high precision. The reader is referred to \cite{tandhi14_2} for a numerical recipe and further details.
\end{remark}

%The expression in \eqref{eq:cp_ia_col1} is less complicated than \eqref{eq:cp_ia}. When the $\sinr$ threshold $T$ is not large, the $\Psi\left(s\,(T-z)^+,tz,M,\alpha\right)$ term can be further simplified as shown next.

%\begin{corollary}[Small-$T$ Approximation]\label{col:cp_ia_col2}
%  For small $T$ the following approximation becomes tight
%   \begin{IEEEeqnarray}{rCl}
%      &&1+\Psi\left(\frac{s\,(T-z)^+}{\hat M_k},\frac{tz}{\hat M_k},M_k,\alpha_k\right)\simeq\hypergeom{-\frac{2}{\alpha_k}}{M_k}{1-\frac{2}{\alpha_k}}{-\frac{s\,(T-z)^++tz}{\hat M_k}}.\IEEEeqnarraynumspace\label{eq:cp_ia_col2}
%   \end{IEEEeqnarray}
%\end{corollary}

%The right-hand side of \eqref{eq:cp_ia_col2} may be easier to evaluate than the original expression since the Gaussian hypergeometric function is available in most numerical software programs. Moreover, its higher-order derivatives with respect to both $s$ and $t$ appearing in \eqref{eq:cp_ia} can be evaluated fairly easily following the same procedure as in \eqref{eq:diff} for differentiating composite functions. %Note that the approximation in Corollary~\ref{col:cp_ia_col2} transforms the interference related part in \eqref{eq:cp_ia} into a form similar to \eqref{eq:int_lap} in Lemma~\ref{lem:int_lap}.

% \begin{remark}[Single-Tier, Single-Tx-Antenna]
%    Setting $K=1$ and $M=1$, we recover the coverage probability result from\cite{tanbourgi14_1} for single-tier single-Tx-antenna cellular networks.
% \end{remark}

\section{Numerical Examples and Design Insights}\label{sec:design}

We next leverage the theoretical results developed in the prior sections to study the system performance through numerical examples. Besides, the approximations introduced in Sections~\ref{sec:int_blind} and \ref{sec:int_aware} are verified by numerical simulations. Unless stated otherwise, we assume $K=3$ with the typical tier-specific system parameters given in Table~\ref{tab:param}, see \cite{3gpp_tr_36814}. The dispersion matrices $\mathbf{A}_{nm}$, $\mathbf{B}_{nm}$ are taken from\cite[Sec.~2.2.3]{shah00}. For the simulation, in each tier an average number of 100 BSs was dropped within a disc of radius $\sqrt{\tfrac{100}{\lambda_k\pi}}$ for a total number of $2\times 10^3$ iterations with the typical receiver assumed in the disc center. In each iteration, a frame spanning $80$ (frequency) resources, each consisting of one or more OSTBC words (depending on the ratio $L_\ell/L_k$), was generated at every BS and the received signal was recorded at the typical receiver. For IA-MRC, the interference in one frame was estimated by averaging over the squared sum envelope of the (possibly non-aligned) received interfering codewords across the $80$ resources according to \eqref{eq:int_gen}. Finally, the linear combining according to \eqref{eq:lin_com_ib}, respectively \eqref{eq:lin_com_ia}, for the two types of MRC was performed and the CDF of the resulting {\it exact} post-combiner $\sinr$ was estimated.

\begin{table}[t]
\renewcommand{\arraystretch}{1.3}
\caption{System Parameters used for Numerical Examples}%\vspace{-.2cm}
\label{tab:param}
\centering
\begin{tabular}{|c||c|c|c|}
\hline
Parameter	&	Tier 1 & Tier 2	&	Tier 3\\
\hline
\hline
BS density $\lambda_k$	& $4\,\text{BS/km}^2$ &$16\,\text{BS/km}^2$ &$40\,\text{BS/km}^2$\\
\hline
BS power $P_k$ & $46\,\text{dBm}$	& $30\,\text{dBm}$ &$24\,\text{dBm}$\\
\hline
BS Tx antennas	$M_k$ & 4	& 2 (Alamouti)	&	1 (no OSTBC)\\
\hline
Path loss exponent $\alpha_k$	&	3.76 & 3.67 & 3.5\\
\hline 
\end{tabular}%\vspace{-.38cm}
\end{table}

\subsection{Multi-Tier \& MIMO: IB-MRC vs. IA-MRC}\label{sec:mimo_mrc}

\begin{figure*}[!t]
	\centerline{\subfloat[Coverage probability IB-MRC]{\includegraphics[width=0.48\textwidth]{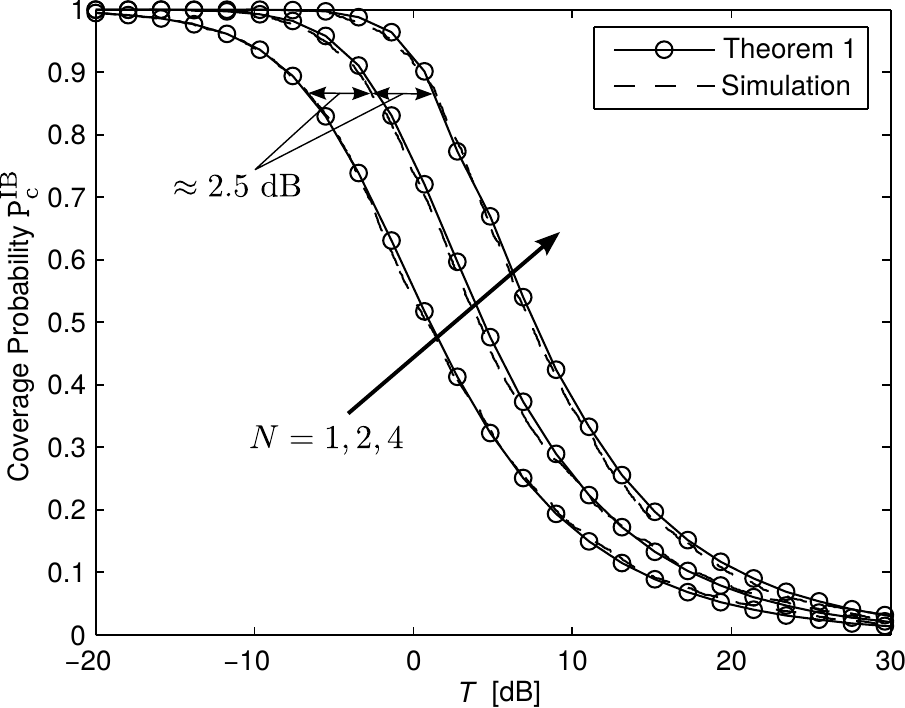}
	\label{fig:cp_ib}}
	\hfil
	\subfloat[Coverage probability IA-MRC]{\includegraphics[width=0.48\textwidth]{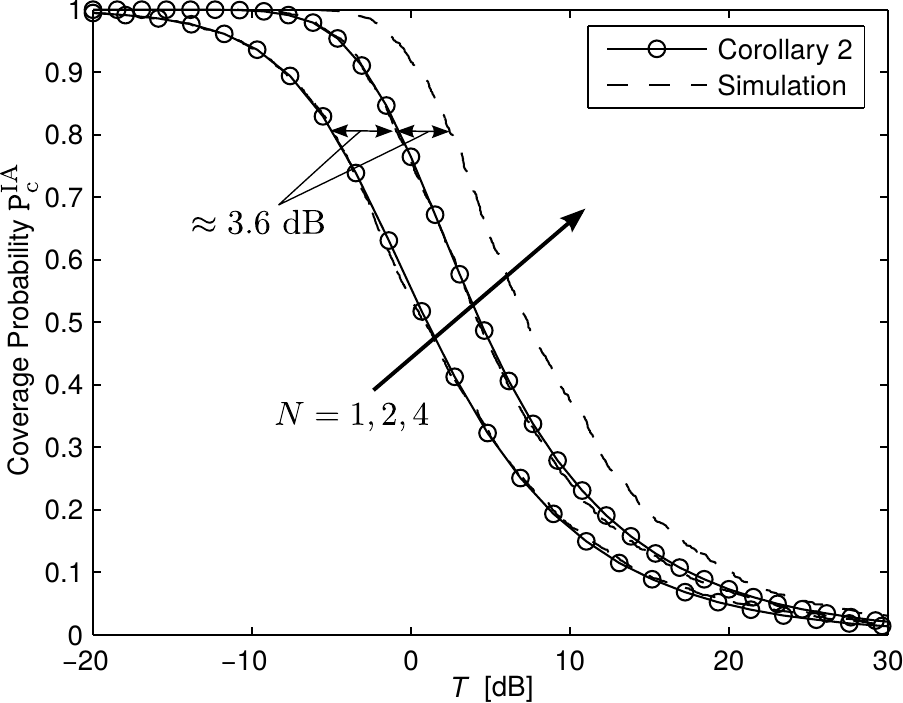}
	\label{fig:cp_ia}}}
	\caption{(a) Coverage probability $\cpib$ for different $N$. Rx noise is $\sigma^2=-104\,\text{dBm}$. (b) Coverage probability $\cpia$ for $\alpha=3.7$, $\sigma^2=0$, and $M_k\equiv M=2$.}\vspace{-.3cm}
\end{figure*}
\begin{figure*}[!t]
	\centerline{\subfloat[Relative coverage probability gain $\Delta^{\text{IA-MRC}}_{\text{IB-MRC}}$]{\includegraphics[width=0.48\textwidth]{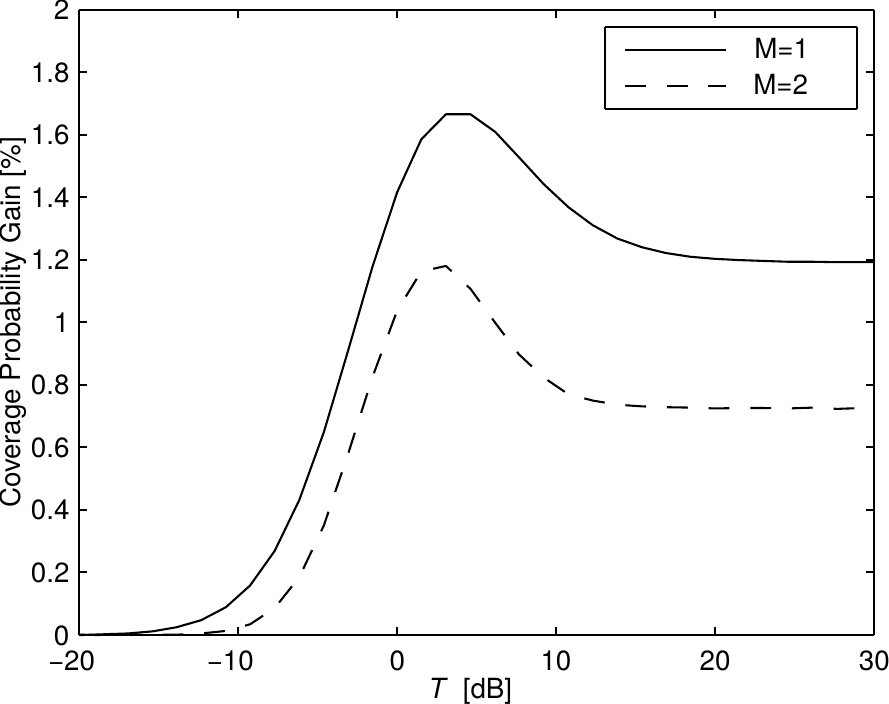}
	\label{fig:cp_comp}}
	\hfil
	\subfloat[Coverage probability of MISO]{\includegraphics[width=0.48\textwidth]{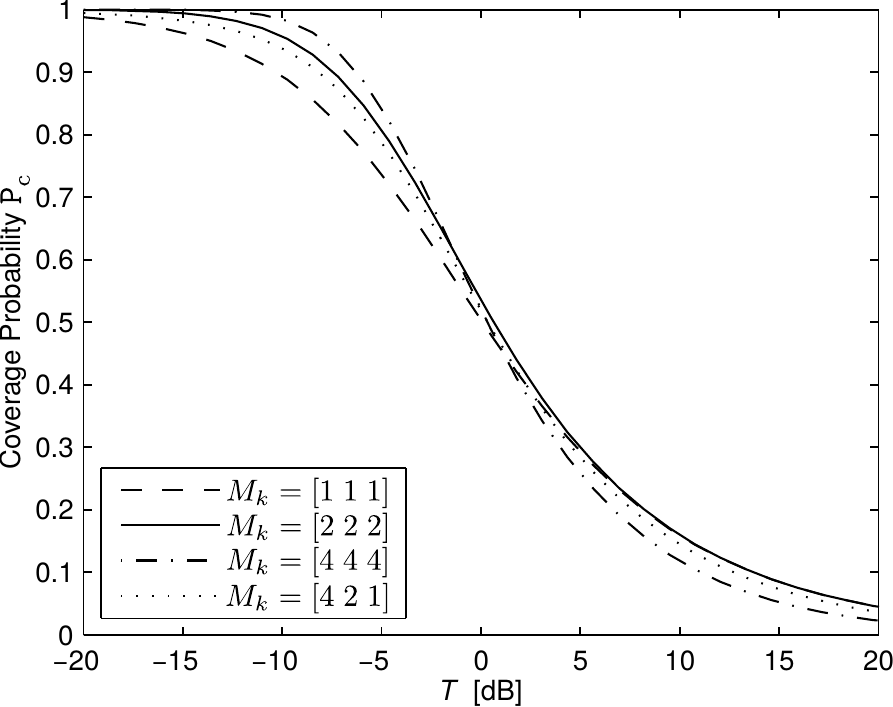}
	\label{fig:cp_miso}}}
	\caption{(a) Relative coverage probability gain $\Delta^{\text{IA-MRC}}_{\text{IB-MRC}}$ for $\alpha=3.7$, $\sigma^2=0$, and $M=1,2$. (b) Coverage probability for MISO case with different antenna configurations. Rx noise is $\sigma^2=-104\,\text{dBm}$.}
\end{figure*}
\setlength{\belowcaptionskip}{-0pt}

We first focus on IB-MRC and consider a typical HetNet scenario with the parameters shown in Table~\ref{tab:param}. The $(4,4,3/4)$-OSTBC from\cite[7.4.10]{larson08} is chosen for tier one. Fig.~\ref{fig:cp_ib} shows the coverage probability $\cpib$ versus the $\sinr$ threshold $T$ for IB-MRC and different number of Rx antennas $N$. It can be seen that the theoretical expressions perfectly match the simulation results, thereby validating the Gamma approximation explained in Section~\ref{sec:int_blind}. As expected, increasing $N$ improves $\cpib$ since the typical user enjoys a larger array gain. For practical target coverage probabilities, i.e. around 80\% covered users, the horizontal gap between the $\cpib$ curves is roughly 2.5~dB. Fig.~\ref{fig:cp_ia} shows the coverage probability $\cpia$ for IA-MRC. Here, we consider the interference-limited case ($\sigma^2=0$) with equal path loss exponents ($\alpha_k=3.7$) and $M_k=2$ for all $k$. Again, simulation results and theoretical expressions are fairly accurate over the entire range of $T$, which justifies the approximation made in Section~\ref{sec:int_aware}, see Remark~\ref{rem:sinr_approx}. For IA-MRC, doubling $N$ yields a gain of around 3.6~dB for the same target coverage probability. %This gain, about 4\% in the practically relevant regime, is comparable to the gain due to frequency-diversity based resource allocation in SISO HetNets\cite{tanbourgi14_2}. The same behavior (gain around 4\%) was observed also for IB-MRC. This rather minor improvement from Tx diversity is well-known in the literature\cite{goldsmith05} and is reconfirmed and quantified for MIMO HetNets with MRC here.

Next, we compare the performance of IB-MRC and IA-MRC for the same scenario, i.e., $\sigma^2=0$, $\alpha_k=3.7$ and $M_k\equiv M$. In Fig.~\ref{fig:cp_comp}, the relative coverage probability gain of IA-MRC over IB-MRC, which is defined as $\Delta^{\text{IA-MRC}}_{\text{IB-MRC}}\mathdef \cpia/\cpib-1$, is shown for $M=1,\,2$ Tx antennas. This relative gain is somewhat disappointing small ($<2\%$ in this example). In fact, IA-MRC becomes even less favorable when adding more Tx antennas. This is due to the fact that adding more Tx antennas effectively smoothes out the fading on the interfering channels, which renders the interference increasingly similar across Rx antennas. This trend was already predicted in Section~\ref{sec:int_stats}, where the second-order statistics of the interference were analyzed. Thus, with almost equal interference across Rx antennas, the performance of IB-MRC and IA-MRC become similar due to less interference diversity. In conclusion, the additional though not overwhelming complexity of IA-MRC must hence be traded-off against an insignificant improvement relative to IB-MRC; with Tx diversity activated, IB-MRC may then be the better choice.

% \begin{figure}[t]
%    \centering
% 	\includegraphics[width=.46\textwidth]{../figures/cp_comp}
% 	\caption{Relative coverage probability gain $\Delta^{\text{IA-MRC}}_{\text{IB-MRC}}$ for $\alpha=3.7$, $\sigma^2=0$, and $M=1,2$.}
% 	\label{fig:cp_comp}
% 	\vspace{-.4cm}
% \end{figure}

\begin{figure*}[!t]
	\centerline{\subfloat[Relative Coverage Probability Gain $\Delta^{\text{IB-MRC}}_{\text{SISO}}$]{\includegraphics[width=0.49\textwidth]{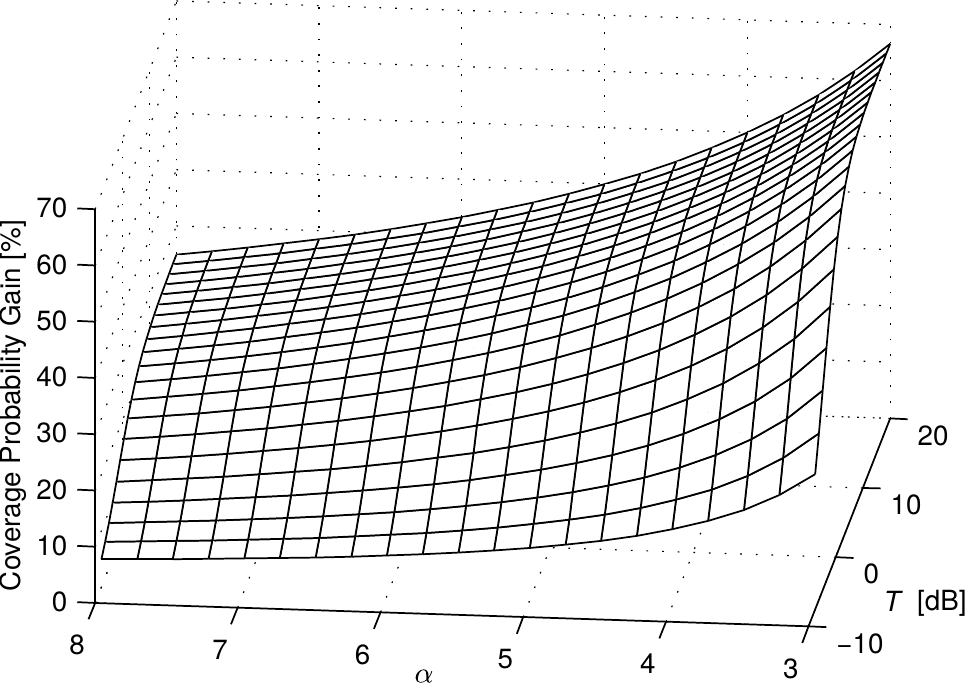}
	\label{fig:gain_ib_siso}}
	\hfil
	\subfloat[Additional Gain $\Delta^{\text{IA-MRC}}_{\text{SISO}}-\Delta^{\text{IB-MRC}}_{\text{SISO}}$ due to IA-MRC]{\includegraphics[width=0.49\textwidth]{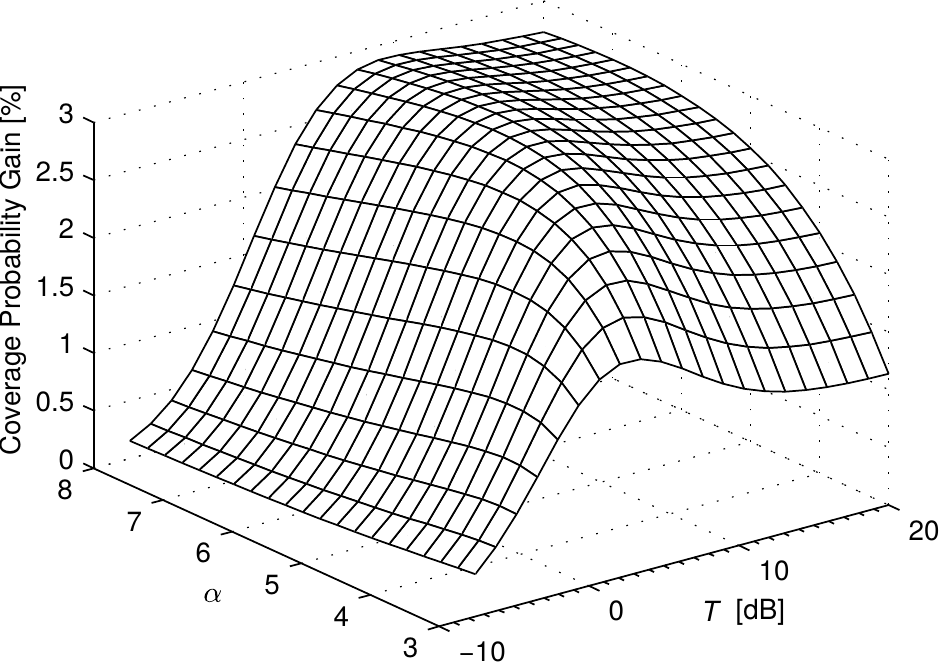}
	\label{fig:gain_ia_ib}}}
	\caption{Parameter: $M=1$, $N=2$, $\alpha_k\equiv\alpha$, and $\sigma^2=0$. (a) Gain of IB-MRC over SISO. (b) Gain of IA-MRC over IB-MRC.}
	%\vspace{-0.4cm}
\end{figure*}

\subsection{Multi-Tier \& MISO: Effect of OSTBC}\label{sec:miso}

As can be inferred from \eqref{eq:cp_ib} in Theorem~\ref{thm:cp_ib}, Tx diversity increases the number of diversity paths on the one hand, while it affects the interference statistics on the other. We next study this trade-off by focusing on OSTBC Tx diversity in a MISO setting ($N=1$) with different Tx antenna configurations. To capture the {\it net gain} of OSTBC one has to account for the rate loss resulting from code rates $r_\ell<1$ when being associated with the $\ell\th$ tier. This can be realized by introducing a tier-specific $\sinr$ threshold $T\Rightarrow T_\ell$ in \eqref{eq:cp_ia} with $T_\ell\mathdef(1+T)^{1/r_\ell}-1$, assuming the Shannon capacity formula $r_\ell\log_2(1+T_\ell)$. The latter adaptation makes sure that the same amount of information is transmitted in every OSTBC. Of course, the same adaptation applies also to the MIMO case with $N>1$. Fig.~\ref{fig:cp_miso} shows that the MISO coverage probability increases only slightly with the number of Tx antennas in the low $\sinr$ regime. For target $\sinr$ larger than a few dB, Tx diversity is not beneficial. In fact, for OSTBCs with rates $r_k<1$, e.g., $M_k=4$, Tx diversity even reduces the coverage probability in this regime due to the aforementioned rate loss. This is in line with prior findings for the single-user case \cite{tarokh99_2} and single-tier cellular networks \cite{direnzo15}, where little to no gains of OSTBC Tx diversity with more than two antennas were reported for reasonable operating points.

\subsection{Multi-Tier \& SIMO: Gain of MRC over SISO}\label{sec:mrc_simo}

From an information-theoretic viewpoint, Rx diversity with MRC is more appealing than Tx diversity with space-time coding as the latter incurs a power penalty\cite{goldsmith05}. But the potential gains of Rx diversity with MRC in HetNets, however, are not yet well-understood due to their dependence on many system parameters. While in Section~\ref{sec:mimo_mrc} the relative performance of IB-MRC and IA-MRC was studied, we now focus on the gains provided solely by MRC Rx diversity ($M_k=1$) over SISO transmission. For that, we consider the interference-limited regime, equal path loss exponents across tiers, and $N=2$. In this case, we obtain the simple expressions
\begin{IEEEeqnarray}{rCl}
	\cpib&=&\frac{1}{\hypergeom{-\frac{2}{\alpha}}{1}{1-\frac{2}{\alpha}}{-T}}\IEEEnonumber\\
	&&\qquad+\underbrace{\frac{\frac{\mathrm d}{\mathrm ds}\hspace{-.05cm}\left[\hypergeom{-\frac{2}{\alpha}}{1}{1-\frac{2}{\alpha}}{-sT}\right]_{s=1}}{\hypergeom{-\frac{2}{\alpha}}{1}{1-\frac{2}{\alpha}}{-T}^2}}_{\mathdef G^{\text{IB}}(\alpha,T)} \IEEEeqnarraynumspace\label{eq:cp_ib_gain}
\end{IEEEeqnarray}
for IB-MRC and
\begin{IEEEeqnarray}{rCl}
    \cpia&=&\frac{1}{\hypergeom{-\tfrac{2}{\alpha}}{1}{1-\tfrac{2}{\alpha}}{-T}}\IEEEnonumber\\
	&&\qquad+\underbrace{\int_{0}^{T}\frac{\tfrac{\mathrm d}{\mathrm dt}\left[\mathcal{A}(T-z,tz,\alpha)\right]_{t=1}}{z\,\mathcal{A}(T-z,z,\alpha)^2}\,\mathrm dz}_{\mathdef G^{\text{IA}}(\alpha,T)}.
   \IEEEeqnarraynumspace\label{eq:cp_ia_gain}
\end{IEEEeqnarray}
for IA-MRC, where $\mathcal{A}(\cdot,\cdot,\cdot)\mathdef1+\Psi(\cdot,\cdot,1,\cdot)$ is
\begin{IEEEeqnarray}{rCl}
	\mathcal{A}(a_1,a_2,q)&=& \frac{a_1}{a_1-a_1}\,\hypergeom{-\tfrac{2}{q}}{1}{1-\tfrac{2}{q}}{-a_1}\IEEEnonumber\\
	&&\qquad-\frac{a_2}{a_1-a_2}\,\hypergeom{-\tfrac{2}{q}}{1}{1-\tfrac{2}{q}}{-a_2}.\IEEEeqnarraynumspace\label{eq:cp_ia_A}
\end{IEEEeqnarray}	

\begin{remark}\label{rem:comp_single_antenna}
The first terms in \eqref{eq:cp_ib_gain} and \eqref{eq:cp_ia_gain} correspond to the SISO coverage probability\cite{jo12}
\begin{IEEEeqnarray}{rCl}
   \cpsiso=\frac{1}{\hypergeom{-\frac{2}{\alpha}}{1}{1-\frac{2}{\alpha}}{-T}}.\IEEEeqnarraynumspace
\end{IEEEeqnarray}
This, in turn, means that $G^{\text{IB}}(\alpha,T)$ in \eqref{eq:cp_ib_gain} and $G^{\text{IA}}(\alpha,T)$ in \eqref{eq:cp_ia_gain} quantify the absolute coverage probability increase of dual-antenna IB-MRC and IA-MRC, respectively, over SISO in HetNets.
\end{remark}

\begin{figure*}[!t]
	\centerline{\subfloat[Deviation $\delta_{\text{NC}}$ and $\delta_{\text{FC}}$ for No-/Full-Correlation Model]{\includegraphics[width=0.475\textwidth]{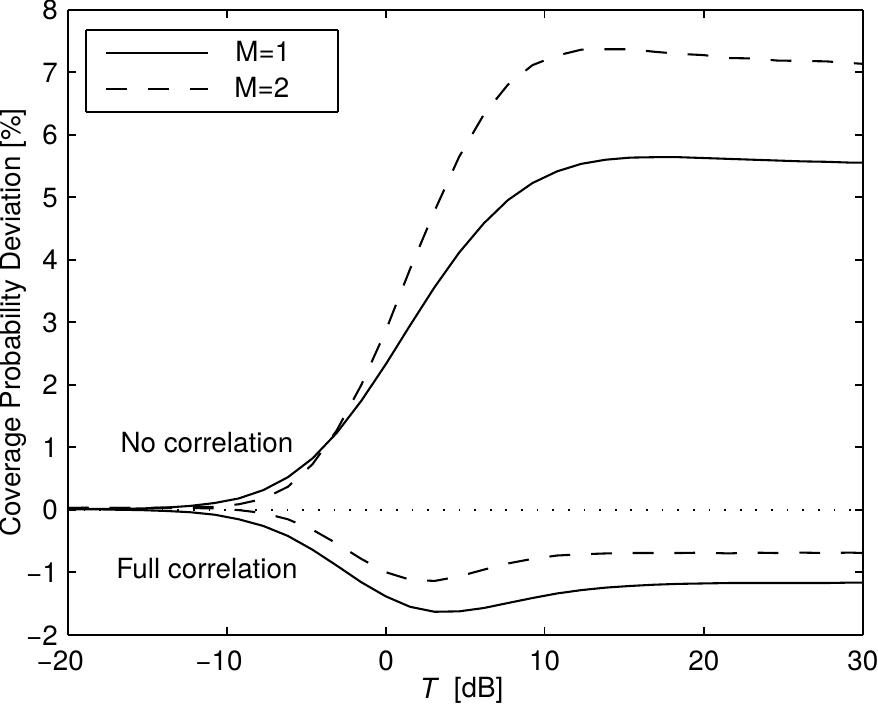}
	\label{fig:cp_gap}}
	\hfil
	\subfloat[Asymptotic Outage Probability for IA-MRC]{\includegraphics[width=0.491\textwidth]{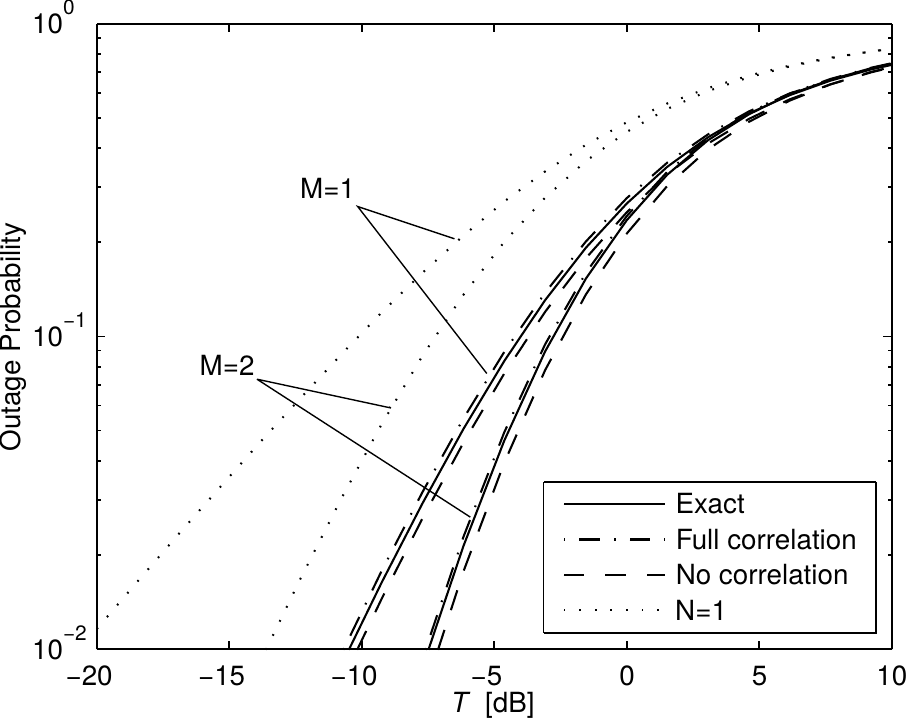}
	\label{fig:cp_div}}}
	\caption{(a) Relative coverage probability deviation $\delta_{\text{NC}}$ and $\delta_{\text{FC}}$ for $M_k\equiv M=1,\,2$. (b) Outage probability for exact correlation, no-correlation and full-correlation model for different $M_k\equiv M=1,\,2$. %Results for single-Rx-antenna case ($N=1$) are also shown for reference. 
	Parameters are $\sigma^2=0$ and $\alpha_k\equiv\alpha=3.7$.}
	\vspace*{4pt}
\end{figure*}

\begin{figure*}[!t]
\normalsize
\setcounter{equation}{33}
\hrulefill
\begin{IEEEeqnarray}{rCl}
	G^{\text{IA}}(\alpha,T)=\int_{0}^{T} \frac{\frac{2T-4z}{z+1}-\left[\left(T\,(2+\alpha)-z\,(4+\alpha)\right)\hypergeom{1}{-\tfrac{2}{\alpha}}{1-\tfrac{2}{\alpha}}{-z}+\alpha\,(z-T)\hypergeom{1}{-\tfrac{2}{\alpha}}{1-\tfrac{2}{\alpha}}{-T+z}\right]}{\alpha\left[z\,\hypergeom{1}{-\tfrac{2}{\alpha}}{1-\tfrac{2}{\alpha}}{-z}+(-T+z)\,\hypergeom{1}{-\tfrac{2}{\alpha}}{1-\tfrac{2}{\alpha}}{-T+z}\right]^2}\,\mathrm dz\IEEEeqnarraynumspace\label{eq:cp_ia_gain2}
\end{IEEEeqnarray}
\setcounter{mycounter1}{\value{equation}}
\hrulefill
\setcounter{equation}{35}
\begin{IEEEeqnarray}{rCl}	   \cpsc&=&2\pi\sum\limits_{\ell=1}^{K}\sum\limits_{n=1}^{N}(-1)^{n+1}\binom{N}{n}\lambda_{\ell}\int_{0}^{\infty}y\exp\left(-\frac{nT}{\snr_\ell(y)}-\pi\sum\limits_{k=1}^{K}\lambda_k \hat P_k^{2/\alpha_k}y^{2/\hat \alpha_k}\hypergeom{-\frac{2}{\alpha_k}}{n}{1-\frac{2}{\alpha_k}}{-T}\right)\mathrm dy\IEEEeqnarraynumspace\label{eq:cp_sc}
\end{IEEEeqnarray}
\setcounter{mycounter2}{\value{equation}}
\hrulefill
\vspace*{4pt}
\end{figure*}
\setcounter{equation}{\value{mycounter1}}

The derivatives inside $G^{\text{IB}}(\alpha,T)$ and $G^{\text{IA}}(\alpha,T)$ can be obtained with the help of \eqref{eq:diff}. For IA-MRC, $G^{\text{IA}}(\alpha,T)$ can be expressed by the integral in \eqref{eq:cp_ia_gain2} at the top of the next page. As a consequence of Remark~\ref{rem:comp_single_antenna}, we can identify $G^{\text{IB}}(\alpha,T)$ and $G^{\text{IA}}(\alpha,T)$ as the characteristic terms for analyzing the gain of dual-antenna MRC relative to SISO transmission. This relative gain can be defined as $\Delta^{\text{IB-MRC}}_{\text{SISO}}\mathdef \cpib/\cpsiso-1$ $=G^{\text{IB}}(\alpha,T)$$\hypergeom{-2/\alpha}{1}{1-2/\alpha}{-T}$, respectively $\Delta^{\text{IA-MRC}}_{\text{SISO}}$ $\mathdef$$\cpia/\cpsiso-1=G^{\text{IA}}(\alpha,T)\hypergeom{-2/\alpha}{1}{1-2/\alpha}{-T}$, for the two MRC schemes.

Figure~\ref{fig:gain_ib_siso} shows the relative coverage probability gain $\Delta^{\text{IB-MRC}}_{\text{SISO}}$ vs. $\alpha$ and $T$. The relative gain monotonically decreases with $\alpha$ and monotonically increases with $T$. Interestingly, $\Delta^{\text{IB-MRC}}_{\text{SISO}}$ saturates at large $T$, although in the corresponding interference-free case the relative gain of MRC over SISO grows unboundedly in $T$ \cite[7.2.4]{goldsmith05}. For typical values $3<\alpha<5$ and $T>-6\,\text{dB}$, the relative gain of IB-MRC is between 12\%--66\%. Fig.~\ref{fig:gain_ia_ib} illustrates the {\it additional} relative gain $\Delta^{\text{IA-MRC}}_{\text{SISO}}-\Delta^{\text{IB-MRC}}_{\text{SISO}}$ when switching from IB-MRC to IA-MRC. In line with Fig.~\ref{fig:cp_comp}, IB-MRC already harvests most of the gains over SISO transmission as the additional improvement of IA-MRC does not exceed 3\%. Nevertheless, the largest additional improvement for realistic path loss exponents around $\alpha=4$ lies entirely in the practical regime $-5<T<10$ dB.

\subsection{Effect of Spatial Interference Correlation}

As explained in Section~\ref{sec:introduction} and \ref{sec:int_stats}, interference correlation across Rx antennas influences the performance of IA-MRC. Mathematically, this can be seen by noting that the post-combiner $\sinr$ in \eqref{eq:sinr_ia_simple} is a sum of correlated random variables. The difficulty of characterizing the coverage probability for general $N$ is due to the complicated correlation structure inherent to this sum, see Section~\ref{sec:int_aware}. To increase mathematical tractability, two simpler correlation models are thus typically used in the literature: 1) no-correlation model, and 2) full-correlation model. Using the results from Section~\ref{sec:int_aware}, the validity of these models for IA-MRC will be discussed next.

\subsubsection{No-Correlation Model}
A commonly made assumption to maintain analytical tractability is to assume that the $\mathsf{I}_{n}$ in \eqref{eq:sinr_ia_simple} are uncorrelated, i.e., the locations of interfering BSs in $\mathsf{I}_{n}$ originate from separate independent point processes for each Rx antenna $n$. Under this assumption, we obtain the following coverage probability $\cpnc$ for the no-correlation model.

\begin{proposition}[Coverage Probability $\cpnc$] \label{prop:nc}
The coverage probability $\cpnc$ for dual-antenna IA-MRC in the no-correlation model is given by \eqref{eq:cp_ia} with $1+\Psi(\cdot,\cdot,\cdot,\cdot)$ replaced by
\begin{IEEEeqnarray}{rCl}
	   &&\hypergeom{-\frac{2}{\alpha_k}}{M_k}{1-\frac{2}{\alpha_k}}{-\frac{s}{\hat M_k}\,(T-z)^{+}}\IEEEnonumber\\
	&&\qquad+\hypergeom{-\frac{2}{\alpha_k}}{M_k}{1-\frac{2}{\alpha_k}}{-\frac{t}{\hat M_k}z}-1.\IEEEeqnarraynumspace\label{eq:cp_nc_general}
	\end{IEEEeqnarray}	
\end{proposition}
\begin{IEEEproof}
See Appendix~\ref{sec:proof_nc}.
\end{IEEEproof}

By comparing the mathematical form of the expression in \eqref{eq:cp_nc_general} with $1+\Psi(\cdot,\cdot,\cdot,\cdot)$ in \eqref{eq:cp_ia}, the influence of spatial interference correlation becomes apparent: in \eqref{eq:cp_nc_general} the first two terms result in a factorization of the PDFs $f_{\sinr_1}(T-z)$ and $f_{\sinr_2}(z)$ in the integral over $z$, which corresponds to the well-known convolution-type integral for sums of independent random variables\cite{feller71}. A closer look at $1+\Psi(\cdot,\cdot,\cdot,\cdot)$ shows that no such factorization of the joint PDF of $\sinr_1$ and $\sinr_2$ can be made due to their correlation across Rx antennas.

\subsubsection{Full-Correlation Model}
Another frequently used approach in the literature to simplify the analysis is to assume that the $\mathsf{I}_{n}$ are fully correlated, i.e., the fading gains $\mathbf{h}_{i,n}$ yield the same realization across $n$ for all $\x_i\in\Phi^{o}$. Under this assumption, the corresponding coverage probability $\cpfc$ for the full-correlation model can be derived for arbitrary $N$ as shown next.

\begin{proposition}[Coverage Probability $\cpfc$] \label{prop:fc}
The coverage probability $\cpfc$ for $N$-antenna IA-MRC in the full-correlation model is the same as for IB-MRC, see \eqref{eq:cp_ib} in Theorem~\ref{thm:cp_ib}.
\end{proposition}
\begin{IEEEproof}
Since $\mathbf{h}_{i,u}\equiv\mathbf{h}_{i,v}$ for all $u,v\in\{1,\ldots,N\}$ and $\x_i\in\Phi^o$, it follows that $\mathsf{I}_{u}\equiv\mathsf{I}_{v}$ for all $u,v\in\{1,\ldots,N\}$. Then, the $\sinr$ expression for IA-MRC in \eqref{eq:sinr_ia_simple} collapses to \eqref{eq:sinr_ib}.
\end{IEEEproof}

Figure~\ref{fig:cp_gap} shows the relative coverage probability deviation for the two simpler correlation models for $M=1,\,2$. The relative deviation is defined as $\delta_{\text{NC}}\mathdef\cpnc/\cpia-1$ ($\delta_{\text{FC}}\mathdef\cpfc/\cpia-1$) for the respective correlation models. First, it can be seen that both models reflect the true performance at small $T$. For $T>0$ dB, the no-correlation model yields a significantly optimistic performance prediction ($3\%<\delta_{\text{NC}}<8\%$), depending on the number of Tx antennas. In contrast, the full-correlation model slightly underestimates the true performance ($\delta_{\text{FC}}<2$\%). In line with our intuition, adding a second Tx antenna increases $\delta_{\text{NC}}$ due to the smaller interference variance and larger interference correlation across Rx antennas, see Section~\ref{sec:int_stats}. Consequently, $\delta_{\text{FC}}$ decreases in this case. Varying the path loss exponent has the same effect as observed in\cite{tanbourgi14_1} for single-tier single-Tx-antenna cellular networks and is therefore not shown here. The smaller deviation for the full-correlation model was already reported in\cite{tandhi14_2} for Aloha-based ad hoc networks and is reconfirmed in this work for HetNets. Fig.~\ref{fig:cp_div} illustrates the outage probability ($1-\cp$) for the exact, no-correlation, and full-correlation model for $M=1,\,2$. It can be seen that the simpler correlation models preserve the true diversity order for dual-antenna IA-MRC. Interestingly, the diversity order due to IA-MRC (which is equal to $N$) remains unaffected by the interference correlation across Rx antennas. This is in sharp contrast to Aloha-based ad hoc networks, where the no-correlation model fails to recover the true diversity behavior \cite{tandhi14_2}. As expected, the diversity order of IA-MRC for $M=2$ and $N=2$ is $NM=4$. 

In conclusion, the full-correlation model offers a tight approximation to the performance of IA-MRC in HetNets, particularly when BSs employ multiple Tx antennas. This result is congruent with the prior observation that the performance gap between IA-MRC and IB-MRC is not significant and further decreases with the number of Tx antennas.

\subsection{Comparison with Selection Combining}

Complexity constraints may sometimes prohibit the use of MRC and allow only for simpler combining schemes. Importantly, while hardware requirements of combining schemes are independent of the communication environment, their performance obviously is not. To properly balance complexity-performance trade-offs, it is hence important to compare the performance of MRC to other combining schemes using a realistic model. We do this next for the example of SC, which is also widely used and less complex compared to MRC\cite{goldsmith05}. A similar comparison can be found in\cite{tandhi14_2} for Aloha-based ad hoc networks and\cite{tanbourgi14_1} for single-tier networks. We next focus on case $M_k=1$ and leave an extension for possible future work. The next result is a generalization of\cite{haenggi12_1} and gives the coverage probability for SC, which we denote by $\cpsc$.

\begin{figure}[t]
   \centering
	\includegraphics[width=.485\textwidth]{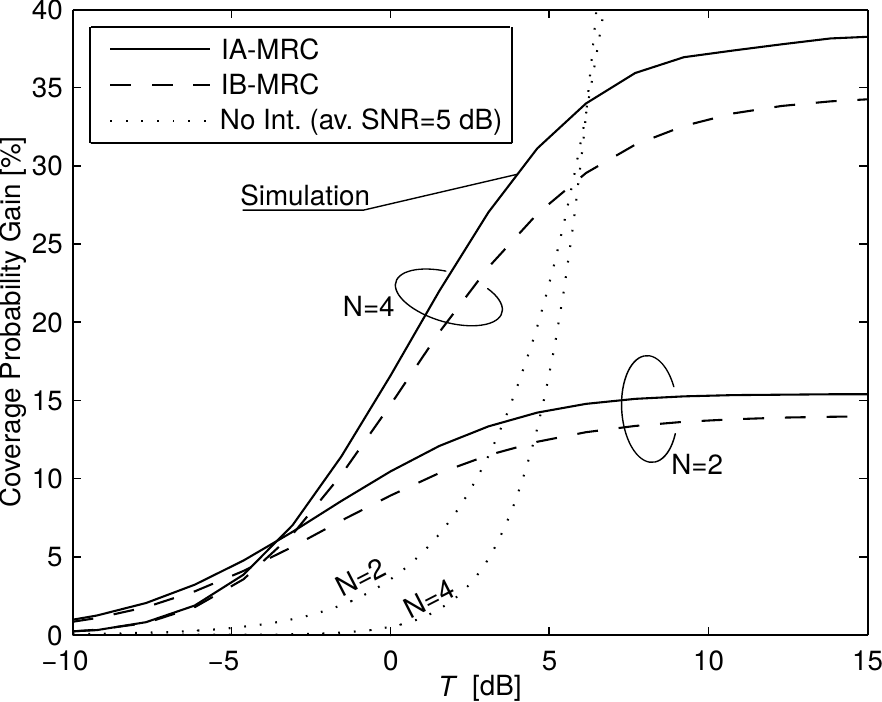}
   \caption{Relative coverage probability gain $\Delta^{\text{IB-MRC}}_{\text{SC}}$ and $\Delta^{\text{IA-MRC}}_{\text{SC}}$ for $N=2,\,4$. Parameters: $\alpha_k\equiv\alpha=3.7$, $M_k\equiv M=1$ and $\sigma^2=0$. Dotted line corresponds to interference-free case and was obtained using \cite[Sec.~7.2]{goldsmith05} with average SNR 5 dB.}
   \label{fig:mrc_sc_gain}
\end{figure}

\begin{theorem}[Coverage Probability $\cpsc$]\label{thm:general_cp_sc}
The coverage probability $\cpsc$ for SC in the described setting with $M_k\equiv M=1$ is given by \eqref{eq:cp_sc} at the top of the last page.
% %{\small
% \begin{IEEEeqnarray}{rCl}	   \cpsc&=&2\pi\sum\limits_{\ell=1}^{K}\sum\limits_{n=1}^{N}(-1)^{n+1}\binom{N}{n}\lambda_{\ell}\int_{0}^{\infty}y\exp\left(-\frac{nT}{\snr_\ell(y)}\right.\IEEEnonumber\\
% &&\left.\hspace{4cm}-\pi\sum\limits_{k=1}^{K}\lambda_k \hat P_k^{2/\alpha_k}y^{2/\hat \alpha_k}\hypergeom{-\frac{2}{\alpha_k}}{n}{1-\frac{2}{\alpha_k}}{-T}\right)\mathrm dy.\IEEEeqnarraynumspace\label{eq:cp_sc}
% \end{IEEEeqnarray}
%}

\setcounter{equation}{\value{mycounter2}}
\end{theorem}
\begin{IEEEproof}
	By\cite[Eq. (8)]{haenggi12_1}, we can express $\cpsc$ as
	\begin{IEEEeqnarray}{rCl}
	   \cpsc&=&\sum\limits_{n=1}^{N}(-1)^{n+1}\binom{N}{n}\mathtt{P}_n(T),
	\end{IEEEeqnarray}
	where $\mathtt{P}_n(T)\mathdef\mathbb{P}(\sinr_1>T,\ldots,\sinr_n>T)$ is the {\it joint success probability}, i.e., the probability of the $\sinr$ being greater than $T$ at $n$ Rx antennas simultaneously. Invoking Lemma~\ref{lem:asso} and following the same line of thoughts as in\cite[Appendix~A]{Haenggi14twc}, the conditional $\mathtt{P}_{n,\ell}(T,y)$ (conditioned on tier $\ell$ and serving BS distance $y$) can be written as
	\begin{IEEEeqnarray}{rCl}
	   &&\mathtt{P}_{n,\ell}(T,y)\IEEEnonumber\\
	   &&\quad\overset{\text{(a)}}{=}\mathbb{E}_{\Phi^o}\left[\prod\limits_{q=1}^{n}\mathbb{P}\left(|\h_{o,q}|^2>\tfrac{Ty^{\alpha}}{P_\ell}(\mathsf{I}_q+\sigma^2)|\Phi^o\right)\right]\IEEEeqnarraynumspace\IEEEnonumber\\
	   &&\quad\overset{\text{(b)}}{=}\exp\left(-\frac{nT}{\snr_\ell(y)}\right)\prod\limits_{k=1}^{K}\mathbb{E}\Big[\prod\limits_{\x_i\in\Phi_k^o}\hspace{-.05cm}\Big(1+T\frac{y^{\alpha_\ell}\hat P_k}{\|\x_i\|^{\alpha_k}}\Big)^{\hspace{-.15cm}-n}\Big]\hspace{-.05cm},\IEEEeqnarraynumspace\label{eq:cp_sc_proof1}
	\end{IEEEeqnarray}
	where (a) follows from the independence of the $\h_{o,q}$ across Rx antennas and (b) follows from the independence of the interferer channel gains $\h_{i,q}$ and from the independence of the $\Phi_k^o$ across $k$. Applying Lemma~\ref{lem:int_lap} and de-conditioning on $\ell$ and $y$ using Lemma~\ref{lem:asso} yields the result.
\end{IEEEproof}

Without Rx noise and with equal path loss exponents across tiers, \eqref{eq:cp_sc} can be further simplified.

\begin{corollary}[Special Case]\label{col:cp_sc}
      In the absence of Rx noise ($\sigma^2=0$) and with equal path loss exponents ($\alpha_k\equiv\alpha$), $\cpsc$ simplifies to
   \begin{IEEEeqnarray}{rCl}
      \cpsc&=&\sum\limits_{n=1}^{N}\frac{(-1)^{n+1}\binom{N}{n}}{\hypergeom{-\frac{2}{\alpha}}{n}{1-\frac{2}{\alpha}}{-T}}.\IEEEeqnarraynumspace\label{eq:cp_sc_col}
   \end{IEEEeqnarray}
\end{corollary}

\begin{remark}[Comment on Corollary~\ref{col:cp_sc}]
This corollary coincides with the result in \cite[Corollary~2]{zhang14} for SC over multiple resource blocks without Rx noise in single-tier networks. Thus, Corollary~\ref{col:cp_sc}, and especially Theorem~\ref{thm:general_cp_sc}, give a generalization of the results from\cite{zhang14} for HetNets.
\end{remark}

Figure~\ref{fig:mrc_sc_gain} shows the relative coverage probability gain of MRC over SC for $N=2,\,4$. The relative gain is defined as $\Delta^{\text{IB-MRC}}_{\text{SC}}\mathdef \cpib/\cpsc-1$ for IB-MRC  and $\Delta^{\text{IA-MRC}}_{\text{SC}}\mathdef \cpia/\cpsc-1$ for IA-MRC. The result for IA-MRC with $N=4$ was obtained by numerical simulations as Theorem~\ref{thm:cp_ia} and Corollary~\ref{col:cp_ia_col1} treat only the case $N=2$. As expected, MRC outperforms SC, particularly at large $T$. However, for practical $T$ around a few dB (which corresponds to between 70-80\% covered users), the gap is less than 10\% for $N=2$. Here, the additional complexity associated with MRC may not be justified as SC achieves similar performance. Nevertheless, adding more Rx antennas increases the relative performance (about 25\% for $N=4$ at practical $T$). Note that, in contrast to the interference-free case (dotted line in Fig.~\ref{fig:mrc_sc_gain}), the relative gain of MRC over SC saturates at large $T$, which is due to the effect of correlated interference.

\section{Conclusion}\label{sec:conclusion}
We developed a tractable model for analyzing downlink MIMO diversity with MRC in HetNets using tools from stochastic geometry. We showed that adding more Tx antennas at the BSs impacts the relative performance of IB-MRC and IA-MRC. One important design insight arising from our analysis is that IA-MRC is less favorable than IB-MRC when OSTBCs for Tx diversity are used. The gains of Tx diversity, however, are visible only at high target coverage probabilities and vanish for $\sinr$ thresholds above a few dB. Another insight is that the full-correlation model yields a tight approximation for IA-MRC; this could enable the analysis of other MIMO techniques, which may be hopeless when considering the exact correlation structure.

Future work may include incorporating other linear combing schemes in combination with Tx diversity in the model. Moreover, a performance comparison between MIMO diversity and more sophisticated MIMO schemes under realistic assumptions, e.g., imperfect CSI and/or imperfect spatial interference cancellation, would contribute to a better understanding of MIMO HetNets. %For instance, this could help finding optimal switching points between different MIMO schemes.

% We presented an analytical framework for analyzing the impact of spatial interference correlation on the downlink performance of dual-antenna MRC receivers in a cellular network. Using tools from stochastic geometry, we derived the coverage probability for a typical dual-branch MRC receiver. Using the theoretical results, we discussed related modeling and design aspects of potential importance to designers of commercial diversity-combining techniques. Future work may include an extension to the case of more than two antennas at the MRC receiver and multiple antennas at the base stations. Another useful direction of future work is to evaluate the downlink rate achievable per user accounting for the load on each base station.

\appendix

%\subsection{Comments on Cell Association}

%Let the spatial configuration of BSs be fixed, i.e., $\Phi=\phi$, and assume that BS $x_1$ from tier $u$ and BS $x_2$ from tier $v$, where $u\neq v$, are received by the typical user with equal average power $P_u\|x_1\|^{-\alpha_u}=P_{v}\|x_2\|^{-\alpha_{v}}=P_{\text{rx}}$. For maximizing coverage, the typical user should then choose the BS providing the larger average SINR as the serving BS. We focus on IB-MRC, however, the same trend is expected to hold also for IA-MRC given the close performance. When BS $x_1$ from tier $u$ is chosen, the average SINR becomes
% \begin{IEEEeqnarray}{rCl}
%    &&\mathbb{E}_{\mathbf{H}_{1},\mathbf{H}_{2},\mathbf{H}_i}\left[\sinr_{u}(\|x_1\|)\right]\IEEEnonumber\\
%    &&\quad=\mathbb{E}_{\mathbf{H}_{2},\mathbf{H}_i}\hspace{-.1cm}\left[\frac{\frac{P_{\text{rx}}}{S_u}\mathbb{E}_{\mathbf{H}_1}\left[\|\mathbf{H}_{1}(M_u)\|_F^2\right]}{ \frac{ P_{\text{rx}}}{S_{v}  }\|\mathbf{H}_2(S_{v})\|_{F}^2+\sum_{k}^{K}\sum_{\x_i\in\Phi^o_k\setminus\{x_2\}}\mathsf{I}_{i,\text{eqv}}+\sigma^2}\right]\IEEEnonumber\IEEEnonumber\\
%    &&\quad=\mathbb{E}_{\mathbf{H}_2,\mathbf{H}_i}\hspace{-.1cm}\left[\frac{M_u/S_u}{\|\mathbf{H}_2(S_v)\|_F^2/S_v+\mathsf{I}_{\text{res}}}\right]\IEEEnonumber\\
%    &&\quad=\frac{M_u}{S_u}\mathbb{E}_{\mathbf{H}_i}\hspace{-.1cm}\left[S_v\mathsf{I}_{\text{res}}e^{S_v\mathsf{I}_{\text{res}}}E_{S_v}(S_v\mathsf{I}_{\text{res}})\right]
% \end{IEEEeqnarray}

\subsection{Proof of Theorem~\ref{thm:cp_ib}}\label{ap:cp_ib}
Applying the law of total probability and making use of Lemma~\ref{lem:asso}, we can express \eqref{eq:cp_general} by
\begin{IEEEeqnarray}{rCl}
   \cp &=& \sum\limits_{\ell=1}^{K}A_{\ell}\int_{0}^{\infty}f_{\mathsf{y},\ell}(y)\,\mathbb{P}\left(\sinr_{\ell}(y)\geq T\right)\,\mathrm dy,\label{eq:cp_proof_ib_step1}
\end{IEEEeqnarray}
where $\mathbb{P}\left(\sinr_{\ell}(y)\geq T\right)$ is the {\it conditional} $\cp$. With Lemma~\ref{lem:fading_dis}, the conditional $\cp$ is written as
\begin{IEEEeqnarray}{rCl}
   &&\mathbb{P}\left(\sinr_{\ell}(y)\geq T\right)\IEEEnonumber\\
   &&\quad=\mathbb{P}\left(\|\mathbf H_o\|_F^2\geq \frac{S_\ell T}{P_\ell y^{-\alpha_\ell}}\left(\sum_{k=1}^{K}\sum_{\x_i\in\Phi^{o}_k}\mathsf{I}_{i,\text{eqv}}+\sigma^2\right)\right)\IEEEnonumber\\
   &&\quad\overset{\text{(a)}}{=}\sum\limits_{m=0}^{NM_\ell-1}\frac{(-1)^m}{m!}\mathbb{E}_{\mathsf{Y}}\left[(-1)^m\mathsf Y^m e^{-\mathsf Y}\right]\IEEEnonumber\\
   &&\quad\overset{\text{(b)}}{=}\sum\limits_{m=0}^{NM_\ell-1}\frac{(-1)^m}{m!}\frac{\mathrm d^m}{\mathrm ds^m}\Big[\mathcal{L}_{\mathsf Y}(s)\Big]_{s=1},\IEEEeqnarraynumspace\label{eq:cp_proof_ib_step2}
\end{IEEEeqnarray}
where in (a) we define $\mathsf{Y}\mathdef\frac{S_\ell T}{P_\ell y^{-\alpha_\ell}}(\sum_{k=1}^{K}\sum_{\x_i\in\Phi^{o}_k}\mathsf{I}_{i,\text{eqv}}+\sigma^2)$ and (b) follows from the differentiation rule for Laplace transforms. With Lemma~\ref{lem:int_lap}, $\mathcal{L}_{\mathsf Y}(s)$ can be obtained as
\begin{IEEEeqnarray}{rCl}
   \mathcal{L}_{\mathsf Y}(s)&=&\exp\bigg(-\frac{sS_\ell T}{\snr_{\ell}(y)}-\pi\sum\limits_{k=1}^{K}\lambda_k\hat P_k^{2/\alpha_k}y^{2/\hat \alpha_k}\IEEEnonumber\\
   &&\qquad\times\left[\hypergeom{-\tfrac{2}{\alpha_k}}{S_k}{1-\tfrac{2}{\alpha_k}}{-\tfrac{s T}{\hat S_k}}-1\right]\bigg),\IEEEeqnarraynumspace\label{eq:cp_proof_ib_step3}
\end{IEEEeqnarray}
where $\snr_{\ell}(y)\mathdef P_\ell y^{-\alpha_\ell}/\sigma^2$ and $\hat S_k\mathdef S_k/S_\ell$. De-conditioning on $y,\ell$ yields the final result.\qed

\subsection{Proof of Theorem~\ref{thm:cp_ia}}\label{ap:cp_ia}
With the law of total probability, Lemma~\ref{lem:asso}, and \eqref{eq:sinr_ia_simple}, we can rewrite \eqref{eq:cp_general} as
\begin{IEEEeqnarray}{rCl}
   \cp &=& \sum\limits_{\ell=1}^{K}A_{\ell}\int_{0}^{\infty}f_{\mathsf{y},\ell}(y)\,\mathbb{P}\left(\sinr_{\ell}(y)\geq T\right)\,\mathrm dy.\label{eq:cp_proof_ia_step1}
\end{IEEEeqnarray}
Next, we focus on $\mathbb{P}\left(\sinr_{\ell}(y)\geq T\right)$, which after conditioning on $\Phi^{o}$, yields
\begin{IEEEeqnarray}{rCl}
	&&\mathbb{E}_{\Phi^{o}}\bigg[\mathbb{P}(\sinr_{1}\geq T-\sinr_{2}|\Phi^{o})\bigg]\IEEEnonumber\\
	&&\quad=\mathbb{E}_{\Phi^{o}}\left[\int_{0}^{\infty}\mathbb{P}(\sinr_{1}\geq T-z|\Phi^{o})\,f_{\sinr_{2}|\Phi^{o}}(z)\,\mathrm dz\right],\IEEEeqnarraynumspace\label{eq:cp_proof_ia_step2}
\end{IEEEeqnarray}
where we have defined the per-antenna conditional $\sinr$
\begin{IEEEeqnarray}{rCl}
   \sinr_{n}\mathdef \frac{P_\ell}{M_\ell y^{\alpha_\ell}}\frac{\|\mathbf{h}_{o,n}\|_{F}^2}{\mathsf{I}_n+\sigma^2}.
\end{IEEEeqnarray}
Applying the same steps as in \eqref{eq:cp_proof_ib_step2}, $ \mathbb{P}(\sinr_{1}\geq T-z|\Phi^{o})$ inside the integral in \eqref{eq:cp_proof_ia_step2} becomes
\begin{IEEEeqnarray}{rCl}
 &&\sum\limits_{m=0}^{M_\ell-1}\frac{(-1)^m}{m!}\,\frac{\mathrm d^m}{\mathrm ds^m}\Bigg[\exp\left(-\frac{sM_\ell ( T-z)^{+}}{\snr_{\ell}(y)}\right)\IEEEnonumber\\
 &&\qquad\times\prod\limits_{k=1}^{K}\prod\limits_{\x_i\in\Phi_k^o}\left(1+\frac{s\,( T-z)^{+}\hat P_k y^{\alpha_\ell}}{\hat M_k\|x_i\|^{\alpha_k}}\right)^{-M_k}\Bigg]_{s=1}.\IEEEeqnarraynumspace\label{eq:cp_proof_ia_step3}
\end{IEEEeqnarray}
Similarly, we have 
\begin{IEEEeqnarray}{rCl}
	f_{\sinr_{2}|\Phi^{o}}(z)&=&\frac{\mathrm d}{\mathrm dw}\big[\mathbb{P}\left(\sinr_2\leq w|\Phi^o\right)\big]_{w=z}\IEEEnonumber\\
	&=&\frac{(-1)^{M_\ell}}{z\,\Gamma(M_\ell)}\,\frac{\mathrm d^{M_\ell}}{\mathrm dt^{M_\ell}}\Bigg[\exp\left(-\frac{t M_\ell z}{\snr_\ell(y)}\right)\IEEEnonumber\\
	&&\times\prod_{k=1}^{K}\prod_{\x_i\in\Phi_{k}^{o}}\Big(1+\frac{tz\hat P_k y^{\alpha_\ell}}{\hat M_k\|x_i\|^{\alpha_k}}\Big)^{-M_k}\Bigg]_{t=1}.\IEEEeqnarraynumspace\label{eq:cp_proof_ia_step4}
\end{IEEEeqnarray}
By Fubini's theorem\cite{feller71}, the expectation $\mathbb{E}_{\Phi^{o}}$ can be moved inside the integral over $z$ in \eqref{eq:cp_proof_ia_step2}. By Leibniz integration rule for infinite integrals\cite{olver10}, the differentiations $\mathrm d^m/\mathrm ds^m$ in \eqref{eq:cp_proof_ia_step3} and $\mathrm d^{M_\ell}/\mathrm dt^{M_\ell}$ in \eqref{eq:cp_proof_ia_step4} can be moved outside $\mathbb{E}_{\Phi^{o}}$. Since the $\Phi_k^{o}$ are independent, we then have
\begin{IEEEeqnarray}{rCl}
   &&\mathbb{E}\left[\prod_{k=1}^{K}\prod_{\x_i\in\Phi_{k}^{o}}\hspace{-.1cm}\left(1+\tfrac{s\,(T-z)^{+}\hat P_k y^{\alpha_\ell}}{\hat M_k\|\x_i\|^{\alpha_k}}\right)^{\hspace{-.1cm}-M_k} \left(1+\tfrac{tz\hat P_k y^{\alpha_\ell}}{\hat M_k\|\x_i\|^{\alpha_k}}\right)^{\hspace{-.1cm}-M_k}\right]\IEEEnonumber\\
   &&\quad=\exp\Bigg\{-\pi\sum\limits_{k=1}^{K}\lambda_k \hat P_k^{2/\alpha_k}y^{2/\hat\alpha_k}\IEEEnonumber\\
   &&\hspace{2.5cm}\times\Psi\left(\frac{s}{\hat M_k}(T-z)^+,\frac{tz}{\hat M_k},M_k,\alpha_k\right)\Bigg\},
   \label{eq:cp_proof_ia_step5}
\end{IEEEeqnarray}
where $\hat M_k\mathdef M_k/M_\ell$ and
\begin{IEEEeqnarray}{rCl}
   \Psi\left(a_1,a_2,p,q\right)=\int_{1}^{\infty}1-\left[\left(1+\frac{a_1}{u^{q/2}}\right)\left(1+\frac{a_2}{u^{q/2}}\right)\right]^{-p}\mathrm du.\IEEEnonumber\\\IEEEeqnarraynumspace\label{eq:cp_proof_ia_step6}
\end{IEEEeqnarray}
Combining \eqref{eq:cp_proof_ia_step1} -- \eqref{eq:cp_proof_ia_step5} yields the result.\qed

\subsection{Proof of Proposition~\ref{prop:nc}}\label{sec:proof_nc}
Recall that in the no-correlation model the interferer locations originate from different point processes, say $\Phi^o$ and $\Phi^{o\prime}$, for each of the two antennas. Then, \eqref{eq:cp_proof_ia_step5} decomposes to
\begin{IEEEeqnarray}{rCl}
   &&\mathbb{E}_{\Phi^o}\left[\prod_{k=1}^{K}\prod_{\x_i\in\Phi_{k}^{o}}\left(1+\frac{s\,(T-z)^{+}\hat P_k y^{\alpha_\ell}}{\hat M_k\|\x_i\|^{\alpha_k}}\right)^{-M_k}\right]\IEEEnonumber\\
   &&\qquad\quad\times\mathbb{E}_{\Phi^{o\prime}}\left[\prod_{k=1}^{K}\prod_{\x_i\in\Phi_{k}^{o}}\left(1+\frac{tz\hat P_k y^{\alpha_\ell}}{\hat M_k\|\x_i\|^{\alpha_k}}\right)^{-M_k}\right]\IEEEnonumber\\
   &&\overset{\text{(a)}}{=}\exp\bigg\{-\pi\sum\limits_{k=1}^{K}\lambda_k \hat P_k^{2/\alpha_k}y^{2/\hat\alpha_k}\IEEEnonumber\\
   &&\hspace{1.5cm}\times\Big[\hypergeom{-\tfrac{2}{\alpha_k}}{M_k}{1-\tfrac{2}{\alpha_k}}{-\tfrac{s}{\hat M_k}\,(T-z)^{+}}-1\IEEEnonumber\\
   &&\hspace{2cm}+\,\hypergeom{-\tfrac{2}{\alpha_k}}{M_k}{1-\tfrac{2}{\alpha_k}}{-\tfrac{t z}{\hat M_k}}-1\Big]\bigg\},\IEEEeqnarraynumspace\label{eq:cp_proof_ia_nc_step1}
\end{IEEEeqnarray}
where (a) follows from Lemma~\ref{lem:int_lap}.
\qed

\end{document}